\documentclass[%
 aip,
 amsmath,amssymb,
 reprint%
]{revtex4-1}

\usepackage{graphicx}% Include figure files
\usepackage{dcolumn}% Align table columns on decimal point
\usepackage{bm}% bold math
\usepackage{xcolor}
\usepackage{changes}
\usepackage[utf8]{inputenc}
\usepackage[T1]{fontenc}
\usepackage{mathptmx}
\usepackage{float}

\definecolor{pink}{rgb}{1,1,0} % color values Red, Green, Blue
\definecolor{red}{rgb}{1,0,0}
\definecolor{yellow}{rgb}{1,1,0}
\definecolor{orange}{rgb}{1,0.5,0}
\definecolor{green}{rgb}{0,1,0}
\definecolor{blue}{rgb}{0,0,1}
\definecolor{white}{rgb}{1,1,1}
\definecolor{purple}{rgb}{0.5,0,0.5}

\begin{document}

\preprint{AIP/123-QED}

%\title[Sample title]{Sample Title:\\with Forced Linebreak}
% Force line breaks with \\
\title[Anatomy of social unrest]{The anatomy of the 2019 Chilean social unrest %\\
%2. The warranty of the Chilean economic model has expired: The anatomy of the 2019 social unrest \\
%3. Economic growth is not enough: The anatomy of the 2019 Chilean social unrest
} %Title of paper

\author{Paulina Caroca}
\affiliation{
 Facultad de Ingenier\'ia y Ciencias, Universidad Adolfo Ib\'a\~nez, \\
 Avda. Diagonal las Torres 2640, Pe\~nalol\'en, Santiago, Chile.
}

\author{Carlos Cartes}
\affiliation{%
 Complex Systems Group, Facultad de Ingenier\'{\i}a y Ciencias Aplicadas, Universidad de los Andes, \\
 Avenida Monse\~nor \'Alvaro del Portillo 12455, Las Condes, Santiago, Chile
}%

\author{Toby P. Davies}
\affiliation{
 Department of Security and Crime Science, University College London, \\
 35 Tavistock Square, London WC1H 9EZ, UK
}

\author{Jocelyn Olivari}
\altaffiliation[Author to whom correspondence should be addressed: ]{jocelyn.olivari@uai.cl}
\affiliation{
 Facultad de Ingenier\'ia y Ciencias, Universidad Adolfo Ib\'a\~nez, \\
 Avda. Diagonal las Torres 2640, Pe\~nalol\'en, Santiago, Chile.
}

\author{Sergio Rica}
\affiliation{
 Facultad de Ingenier\'ia y Ciencias, Universidad Adolfo Ib\'a\~nez, \\
 Avda. Diagonal las Torres 2640, Pe\~nalol\'en, Santiago, Chile.
}

\author{Katia Vogt-Geisse}
\affiliation{
 Facultad de Ingenier\'ia y Ciencias, Universidad Adolfo Ib\'a\~nez, \\
 Avda. Diagonal las Torres 2640, Pe\~nalol\'en, Santiago, Chile.
}

%\author{A. Author}
%\author{B. Author}%
% \email{Second.Author@institution.edu.}
%\affiliation{ 
%Authors' institution and/or address%\\This line break forced with \textbackslash\textbackslash
% \homepage{http://www.Second.institution.edu/~Charlie.Author.}
%\affiliation{%
%Second institution and/or address%\\This line break forced% with \\
%}%

\date{\today}% It is always \today, today,

\begin{abstract}
We analyze the 2019 Chilean social unrest episode, consisting of a sequence of events, through the lens of an epidemic-like model that considers global contagious dynamics.  We adjust the parameters to the Chilean social unrest aggregated public data available from the Undersecretary of Human Rights, and observe that the number of violent events follows a well-defined pattern already observed in various public disorder episodes in other countries since the sixties. Although the epidemic-like models display a single event that reaches a peak followed by an exponential decay, we add standard perturbation schemes that may produce a rich temporal behavior as observed in the 2019 Chilean social turmoil.  Although we only have access to aggregated data, we are still able to fit it to our model quite well, providing interesting insights on social unrest dynamics.
\end{abstract}

\maketitle

\begin{quotation}
The social unrest that has been simmering in Chile for a number of years ended up in October 2019 with a massive turmoil in the capital city Santiago that quickly turned violent and started spreading into other cities within the country, provoking costly damage to public and private infrastructure, and causing serious injuries to a group of the population. Epidemic-like models can be applied to understand the dynamics of complex social phenomena such as social unrest episodes in which contagion effects could take place. Therefore, being able to describe the mechanism and predict the temporal evolution of a potentially violent crowd can provide useful insights to authorities involved in politics and security who want to make informed decisions with a view to controlling and mitigating these episodes of social unrest. As extant literature has already noticed for other countries, the Chilean social unrest episodes depict a standard dynamic in which a single riot event reaches a peak that is then followed by an exponential decay. However, the addition of an external forcing changes dramatically this simple picture in a complex scenario characterized by a sequence of events, like the ones  observed in the Chilean social unrest  episode.
\end{quotation}

\section{Introduction}
\label{Sec:Intr}

Social unrest due to economic, political and/or social factors was all over the world news in 2019. A recent report on global political risk\cite{Verisk} shows that a quarter of all the world's countries saw a significant  upsurge in civil unrest during 2019, including locations as diverse as Chile, Haiti, Hong Kong, Lebanon, Nigeria, Sudan and Venezuela. Furthermore, prior to the Covid-19 pandemia, forecasts showed that this worrying trend would be likely to continue in 2020.

The problem of public disorder, riots and civil unrest has been studied for more than forty years\cite{spilerman1970causes,Burbeck1978,Granovetter,berestycki2015model,Baudains2016london,bonnasse2018epidemiological}. The causes behind each episode are complex and diverse, given the unique institutional framework that prevails in each country. Still, all of them somehow end up disrupting the daily lives of the citizenship and, if mishandled, can sometimes cascade into large-scale violent manifestations of criminality\cite{salehyan2012social}. Depending on the depth of the social turmoil, it can cause substantial damage to transport systems and to private and public property, while also disrupting supply chains, which negatively impacts economic activity and growth. Furthermore, they can also become violent, thus evolving into a public security problem that can provoke serious injuries and death. Even more extremely, if it scales into a nation wide phenomena, it can pave the way for political instability or even a government overthrow, like in Tunisia and Egypt at the aftermath of the Arab Spring in 2011.

Recently, in October 2019 the world saw the \textit{poster child} of Latin America, Chile, go through massive demonstrations in the capital city of Santiago, triggered by rising subway fares. Within a few days, protesters turned violent and the social turmoil rapidly spread to other regions of the country, forcing local authorities to declare a state of emergency and impose military curfews in several cities. The first month of unrest alone caused an estimated USD\$4.6 billion worth of infrastructure damage, and cost the Chilean economy around USD\$3 billion, or 1.1\% of its Gross Domestic Product (GDP) \cite{Verisk}.  

 Being able to describe and predict the temporal evolution of a potentially violent crowd can provide useful insights to authorities involved in politics and security who want to make informed decisions in terms of controlling and mitigating these episodes of social unrest. It is in this context that mathematical modelling and numerical simulations offer a great advantage over empirical observation, as they have the potential to  describe the mechanics and the factors affecting the dynamics of the social event, and hence provide profound understanding  of the social phenomena under study. In particular, such modelling allows distinct scenarios to be explored, as well as the possibility to test different control or mitigation strategies at low cost, and without the ethical concerns of field experiments. Although the control of riots is only one aspect of the problem - disorder has complex long-term sociological causes which also mandate attention - the near-inevitability of periodic social tensions means that mitigation must be considered as part of any strategy.

Epidemic-like models can be applied to social phenomena such as social unrest episodes in which contagion effects could take place. Burbeck, Raine and Abudu Stark introduced this approach in the seventies to analyze large-scale urban riots in Los Angeles, Detroit and Washington D.C. (See Ref. \onlinecite{Burbeck1978}). Inspired by clinical sciences, one initial case of a contagious disease in a large-scale population can easily spread into an epidemic if the infectious individual comes into contact with a sufficiently large number of susceptible individuals, where each contact  - between a susceptible individual with an infectious one - is accompanied by some chance that the susceptible person will contract the infection. In less dense groups, the infected person comes into contact with fewer susceptibles and the chances that he/she will cease to be infectious before successfully infecting someone else are greater \cite{Burbeck1978}. In fact, the authors find that behavioral epidemics can explain the contagion process of riots that occurred during the sixties in United States, which requires susceptibility of individuals to get involved into rioting as well as contact with the riot;  among other things, they observe that that propagation of social unrest does not require the spatial displacement of individual rioters. 

More recently, Bonnasse-Gahot \textit{et al.\@}\cite{bonnasse2018epidemiological} applied similar ideas to the 2005 French riots, which started in a poor suburb in Paris and within three weeks had spread all over France. In line with Burbeck \textit{et al.\@}, they found that although there was little displacement of rioters, the riot activity did travel. On the other hand, Davies \textit{et al.\@}\cite{Davies2013,davies2015event} claim, using a hybrid of epidemiological and transport models, that spatial displacement of individuals is at the basis of the London 2011 episodes of large-scale disorder. The results suggest that the distribution of rioting can be understood in terms of the spatial configuration of London, with highly transited areas at greater risk than others;  this coincides with empirical research that shows, for example, that the proximity of a particular location to underground train stations, or other public transport nodes, increases its probability of being looted \cite{Baudains2016london}. % {\na Even though spatially structured models may capture dynamics between locations, these models may be difficult to study if appropriate data is not available, such as is the case for metapopulation disease models, since they may fail to describe correct human behavior on local scales \cite{wesolowski2016connecting}.}

In this paper we aim to answer the following research question: Can we model the temporal evolution of the 2019 Chilean riot using epidemiological models? To answer this question we compare the available aggregated data recorded by the Undersecretary for Human Rights of the Chilean Ministry of Justice and Human Rights with the solution of such models. We also compare the main traits of Chilean rioting episodes with other similar events around the world.  An interesting feature of the Chilean social unrest episode, regarding its dynamics, is the existence of a number of sizeable rebounds showing particularly rich dynamics not observed in the usual epidemic-like model, meaning that a reformulation is needed  to fully describe the evolution of events observed in  Chile.

The current paper is organized as follows. Section \ref{Sec:Chile} provides some background information on the Chilean economic and social context, and on the social unrest episodes that started in 2019; the empirical behavior of these events is also discussed. Next,  Section
 \ref{Sec:Epidemic} presents an epidemic-like model that considers global variables such as the number of rioting events, rioters, and the susceptible population to join a riot. Some extensions that could improve the explanatory power of the model are presented, such as the addition of an external forcing that can influence the susceptibility of the population to join a rioting episode, which can end up predicting the existence of a temporal behavior that may follow  a periodic, quasi-periodic, or chaotic dynamics. In Section \ref{Sec:DataAnalysis},  the results of the data analysis are presented. Finally, in Section \ref{Sec:Discussion} we provide some final remarks, discuss  implications, and outline new perspectives regarding social unrest episodes; a worrying trend that is likely to continue in the upcoming years.

\section{The 2019 Chilean social unrest}
\label{Sec:Chile}

\subsection{Economic and Social Context}
\label{Sec:Context}
The Chilean economy has been usually praised as a successful case since neoliberal reforms were implemented during the dictatorship period, which lasted from 1973 until early 1990. Since democracy was restored in 1990, the seven ensuing governments continued with the market economic model, although with different ``flavors'' as a result of social and economic reforms, and specific policy approaches implemented by each of them in response to the national and international context~\cite{fd2014}. The Chilean economy went through various phases, some of which were quite remarkable in terms of economic growth and improvement of social indicators:  during the period 1990-98, for example, annual GDP growth rates averaged 7.1\% \cite{fd2014} and the Gini coefficient dropped from 57.2 to 55.5, according to estimates of The World Bank (a Gini index of 0 represents perfect income equality, while an index of 100 implies perfect inequality)~\cite{wbdata}. Some other phases were marked by the country's vulnerability to external turbulences, such as the  Asian crisis in 1998 and the global crisis in late 2008 and 2009. Overall, since 1990 the country's economic performance facilitated the improvement of several social indicators in the upcoming decades, such as the percentage of the population living in poverty (on US\$ 5.5 per day) which dropped from 46.1\% in 1990 to 6.4\% in 2017, while life expectancy at birth rose from 73.5 years old to almost 80 during the same period~\cite{wbdata}. 

% political leaders from both left and right leaning political parties promised Chileans that a free market economy would bring prosperity to everyone. The decades of economic growth (\textbf{add REF}) that followed, together with the implementation of structural reforms and sound macroeconomic policies, 
%

Sound macroeconomic policies and fiscal responsibility eventually allowed Chile to become in 2010 the first South American country to be a member of the Organization for Economic Cooperation and Development (OECD), an exclusive club of thirty seven countries that together account for % 63\% 
almost two thirds of the world's GDP. For years Chile has been the \textit{poster child} of Latin America and has often been considered an example of economic success. After all, steady growth allowed per capita GDP to grow almost sixfold according to The World Bank, from US\$4.5k in 1990 to US\$25.2k in 2018, exceeding the US\$16.6k average per capita income of Latin American and Caribbean countries (per capita GDP is measured in purchase power parity and constant 2011 USD)~\cite{wbdata}.

As previously mentioned, indicators that measure income inequality, such as the Gini coefficient, seem to have been improving in the last decades, which according to the World Bank was reduced from 57.2 in 1990 to 44.4 in 2017 (See Ref. \onlinecite{wbdata}). However, this improvement is mild in relative terms, since Chile remains one of the most unequal countries among OECD economies \cite{flores2019top}. And although over the past three decades, the gap between the rich and poor has widened in the large majority of OECD countries, the average estimated Gini, excluding Chile, is a much lower 31.1 (See Ref. \onlinecite{OECD2020}). Furthermore, according to ECLAC, the richest 1\% of the Chilean population concentrated almost 27\% of the net wealth of the country in 2017 \cite{eclac2019}. These factors may contribute to the widespread feeling of social discontentment and unfairness that simmered in many Chileans in the years leading up to October 2019. In fact, for the past few years Chile has seen an upswing of diverse social movements; the Student Movement, the Mapuche Movement, the Labor Movement, the Feminist Movement, and the Environmental Movement are among the widespread demonstrations that have influenced the political arena \cite{Donoso2017}.\footnote{ Professor Claudia Sanhueza declared to BBC that ``This wave of protests may have been kickstarted by a rise in the price of metro tickets, but the resentment goes back further than that''. Moreover she ``singles out 2006 as a crucial year'' referring to the so-called ``Penguin Revolution'' led by high school students. https://www.bbc.com/news/world-latin-america-50151323.} For example, in 2006 and 2011 the Student Movement led thousands of students into the streets to demand better public education, more social justice and equal opportunities \cite{cabalin2012neoliberal}. It is perhaps, therefore, unsurprising that a share of the Chilean population, especially the rising middle class, claims that after years of unfulfilled promises regarding economic prosperity for all, eventually \textit{``Chile woke up''}, culminating in the massive protests in October 2019.

% that quickly turned violent and started spreading into other cities within the country causing serious damage to public and private infrastructure. The straw that broke the camel's back was. As protests quickly turned violent in the capital city, with masked riots lighting barricades and fighting the police, the turmoil started spreading into other cities within the country. 

\begin{figure}[h]
\begin{center}
  \includegraphics[width=8.5cm]{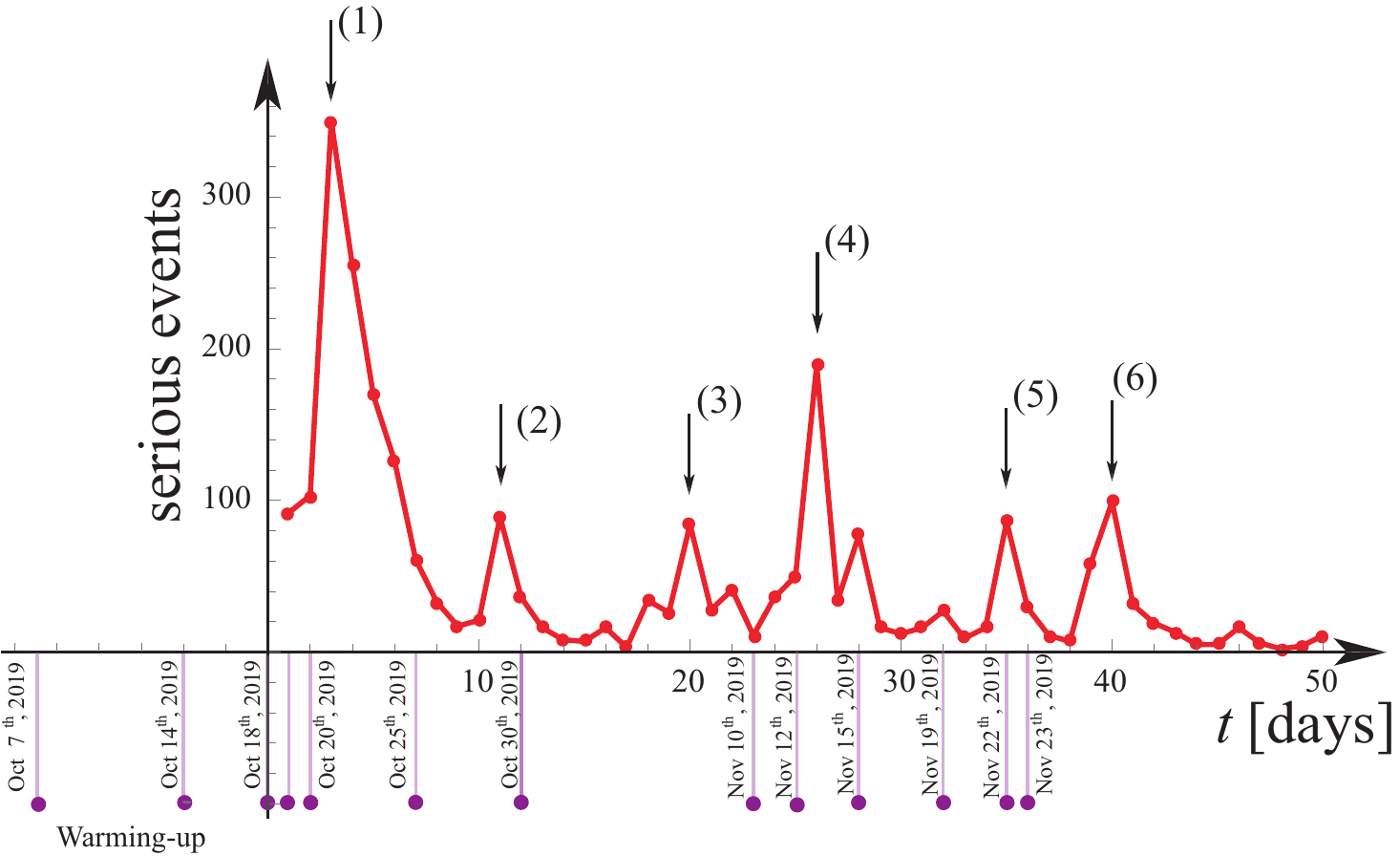}  
 \end{center}
\caption{\label{Fig:TimeLine} Timeline of the 2019 Chilean social crisis. The warm-up phase took place between October 7$^{th}$ and October 18$^{th}$ 2019, when the massive and violent social unrest was triggered. (Data from Ref. \onlinecite{hhrr}).}
\end{figure}

In Fig.\ref{Fig:TimeLine} we summarize the key events that occurred between October and November 2019 (see more detail in Appendix \ref{Sec:timeline}). The protests, led by high school and college students, started on October 7$^{th}$ in the capital city of Santiago due an increase in the subway fare that provoked public outcry. After two weeks, protests quickly turned massive and violent, and the turmoil started spreading into other cities within the country. On October 18$^{\rm th}$, students in Santiago massively evaded the subway fee in defiance over the fare increase. As police attempted to stop students disturbing subway stations, protesters spilled out into the street, burning and destroying metro stations and looting supermarkets, pharmacies, and shops. Since rioters could not be controlled by the local police, the President, Sebasti\'an Pi\~nera, decided to declare a state of emergency to safeguard public order. Military forces were deployed to the streets in many cities and curfews were enforced. After days of both violent and peaceful demonstrations, it was understood that the cause behind the protests was beyond the increase in the subway fare, and it included demands for a substantial improvement in living standards, pensions, health care and education, among other concerns. Furthermore, protesters wanted these improvements right away. After these violent episodes, and a massive peaceful demonstration on October 25$^{th}$ that gathered more than a million people\footnote{The total population in Santiago is about 5.5 million and about 19 million in all Chile.} in different locations in Santiago and the rest of the country, the President eventually yielded to the pressing demands of the population. Since then, some reforms have been announced, including an agreement to scheduled a referendum for April 26$^{th}$ 2020 to vote for a change in the Chilean Constitution written in 1980. However, due to the Covid-19 pandemic, the referendum has been postponed to October 2020. This external shock seems also to have \textit{paused} massive protests due to mandatory and voluntary lockdowns.

 \subsection{Social crisis and catastrophe theory}
\label{Sec:Crisis}

The disorder observed in Chile in 2019 displayed many features in common with previous episodes of a similar nature. In particular, one general signature is the existence of different phases as the disorder evolves: a slow process of \textit{warming up} over an extended period, followed by an abrupt outbreak in which the system rapidly transitions to a peak of disorder, before a relaxation phase during which disorder levels gradually return to a low level. This pattern has been consistently observed for previous cases of rioting, including the aforementioned instances in multiple US cities\cite{Burbeck1978}, Paris\cite{bonnasse2018epidemiological} and London\cite{Davies2013}. In each case, the rapid escalation was seen to coincide with a triggering event: in the majority of those cases, these took the form of instances of apparent police brutality.

While this evidence suggests that such triggering events are a \textit{necessary} ingredient for riots to occur, however, it is clear that they are not in themselves \textit{sufficient}. While the events in question are undoubtedly serious, other events of a similar nature occur frequently without generating riots. This implies that the potential for an event to trigger an outbreak is dependent on more general social conditions; that the riot will only occur if tension is already high. There have been several attempts to explain the mechanics of this process \cite{Epstein7243,pires2017modeling,berestycki2015model}. In general, it is hypothesized that riots occur in situations in which some latent grievance - representing a combination of hardship and absence of legitimacy - accumulates in time. Many factors may contribute to this, including economic factors\cite{flamm2005law}, food scarcity\cite{walton1994free}, racial tensions and the reputation of police. Berestycki \textit{et al.\@}\cite{berestycki2015model} characterize these as contributing to a scenario in which the social system is `ripe' for disorder. In Chile, unfulfilled expectations and a decrease in the legitimacy of the governing class appear to have played this role, reaching a critical mass over a number of years.

More generally, the idea of a transition-like mechanism applied to decisions and crisis in the context of social sciences has been circulating over the last 40 years.  Granovetter\cite{Granovetter} for example, applied threshold models of collective behavior to understand rioting behavior, while Koselleck\cite{Koselleck}, from a more philosophical perspective, extends the concept of crisis into the field of economics. Moreover, as proposed by Beinhocker\cite{beinhocker2006origin}, an economy can be understood as a complex adaptive system of interacting agents that adapt to each other and their environments. In contrast to the views of Traditional Economics, in which the system always converges to an equilibrium, an economy is an open system that sometimes can be stable in an equilibrium-like state, or can exhibit very unpredictable behavior patterns that are far from equilibrium patterns, such as exponential growth, radical collapse or oscillations.

Relatedly, the different phases of social crises that follow a slow evolution that culminate in a fast transition, or a ``catastrophe'', resemble the ideas proposed by the Theory of Catastrophes, developed by Thom\cite{Catastrophe,Zeeman}.  Catastrophe Theory can describe the evolution of a process that depends on a set of control parameters, and whose qualitative behavior can abruptly change under small variations of these parameters (the point at which the ``catastrophe'' happens). Although sometimes this theory has been applied to areas of knowledge such as biology or social sciences in a heuristic fashion (see Zeeman\cite{Zeeman}), the evolution of the state of social tension experienced by a population can also be understood in these terms. Socio-economic factors may increase or decrease this social tension and, eventually, an insignificant variation in some of them ({\it e.g.} a rise in the subway fare), may suddenly increase the social tension in such a way that a ``catastrophe'' occurs. This is interpreted as an abrupt change in the social behavior of the people. Interestingly, because the transition is abrupt (a first order phase transition) it is irreversible: that is, reversing slightly the triggering factor of the ``catastrophe'' does not lead the society back to the original situation.

 Therefore, given that economy can be understood as a complex adaptive system\cite{beinhocker2006origin} and because of the generic features of  catastrophe theory \cite{Catastrophe,Zeeman}, one may claim that the Chilean 2019 social unrest episode has completely changed the \textit{state} in which the country was embedded. In fact, the population has assured that \textit{``Chile woke up''}, and all the measures that are being implemented to control the social crisis are expected to lead the economy towards a \textit{``new normality''}, but not back to the original state.  It is yet to be seen how this social crisis will evolve after the current health emergency, \textit{Covid-19}, starts to fade.

\subsection{Empirical behavior of the social unrest}\label{SubSec:Empirical}
\begin{figure*}[h!]
\begin{center}
 \includegraphics[width=18cm]{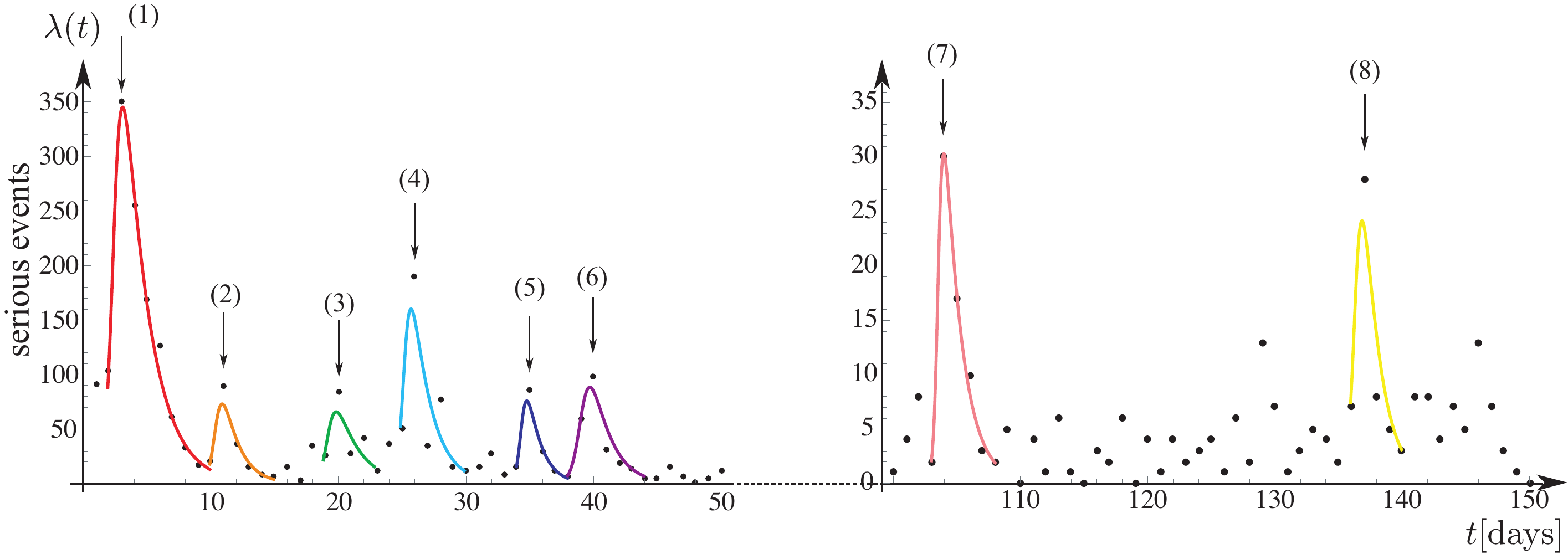} 
 \end{center}
\caption{\label{Fig:DataHR}   The plot of serious events per day are presented as a function of time. $t=0$ is set on October 18$^{th}$ 2019, the day that the violent social unrest started in Santiago downtown. During the first 50 days, we notice the existence of about 6 extreme events. The colored curves represent the theoretical fits using the epidemic model (see Sections \ref{Sec:Epidemic} and \ref{Sec:DataAnalysis}). The fits are labeled as follows: The fit of the \textit{extreme} event (1) corresponds to the red curve; (2) corresponds to orange; (3) corresponds to green; (4) corresponds to cyan; (5) corresponds to blue; and, (6) corresponds to purple. After a quiet period during Christmas 2019, few smaller events can be observed. Events (7) and (8), presented to the right of the figure, correspond to pink and yellow, respectively. We zoom in by 10 times the scale of events (7) and (8).} The fitting parameters will be discussed in Section \ref{Sec:DataAnalysis} and listed in the Table \ref{Table}. 
\end{figure*}

The Undersecretary for Human Rights of the Chilean Ministry of Justice and Human Rights releases daily reports with summarized data on the number of \textit{serious events} per day at the country level~\cite{hhrr}. The data source, provided by the Ministry of Interior and Public Security, comes from the Chilean Police. Serious events include looting, fire, destruction of private or public property and other events of the same nature~\cite{hhrr}. Since the report is based on integrated data, it does not include key information such as the category and magnitude of these serious events, or the location where they took place. This prevents us from analyzing key dimensions of the phenomenon, like contagion effects between cities. Nevertheless, the data is rich and some interesting patterns can be observed from the aggregated data, as illustrated in Fig. \ref{Fig:DataHR}, where $\lambda(t)$ represents the number of serious events that occur at a given time $t$.  Next we describe the behavior of these events.

%\begin{widetext}
 %\end{widetext}

\begin{enumerate}
\item The most prominent episode was event (1), which started in October 18$^{th}$ 2019 in Santiago, reaching 350 serious events by {October} 21$^{\rm st}$ after it started spreading to other cities in the country, as discussed in Section \ref{Sec:Context}. 

\item Events (3) and (4) in Fig. \ref{Fig:DataHR} illustrate the limitations of aggregated data, since these peaks may correspond to spatially separated riot episodes occurring in different cities.  %Because of this, we are not able to delineate each local riot episode and study contagion effects between them.

\item Another key observation is that serious events occurred rather periodically, on a weekly basis. This is possibly explained by the fact that most violent acts occurred on Friday afternoons, with these becoming somehow a symbolic day of the social movement. 

\item  
Although our analysis is based on the available data that is of an aggregated nature, we hypothesize that Chilean events could satisfy a scale-independent behavior as observed in Ref. \onlinecite{bonnasse2018epidemiological} for the case of France,  and which may be influenced by social media use, as has been participation in Chilean protests before\cite{scherman2015student}. Therefore, based on this assumption, every major event is expected to be ensued by a dissipation process that decreases exponentially in time, with an event mean lifetime of the order of one up to two days. 
Prior evidence\cite{Burbeck1978} shows that the mean lifetime of an event was about eight hours in the Los Angeles 1965 social turmoil; about twenty hours in Detroit 1967; about twelve hours in Washington D.C. 1968;  and up to 4 days in the case of French riots in 2005\cite{bonnasse2018epidemiological}. The exponentially decreasing behavior we observe for the Chilean case is depicted in Fig. \ref{Fig:DataHRSlopes}, which plots the same data as in Fig. \ref{Fig:DataHR} but using a semi-logarithmic scale. Under this scale, the linear behavior in time corresponds to an exponential behavior of the original variable. As far as we can see, the first six events follow an approximately linear behavior after each event was triggered. The mean lifetime is inversely proportional to the corresponding slope.

\begin{figure}[h!]
\begin{center}
 \includegraphics[width=7cm]{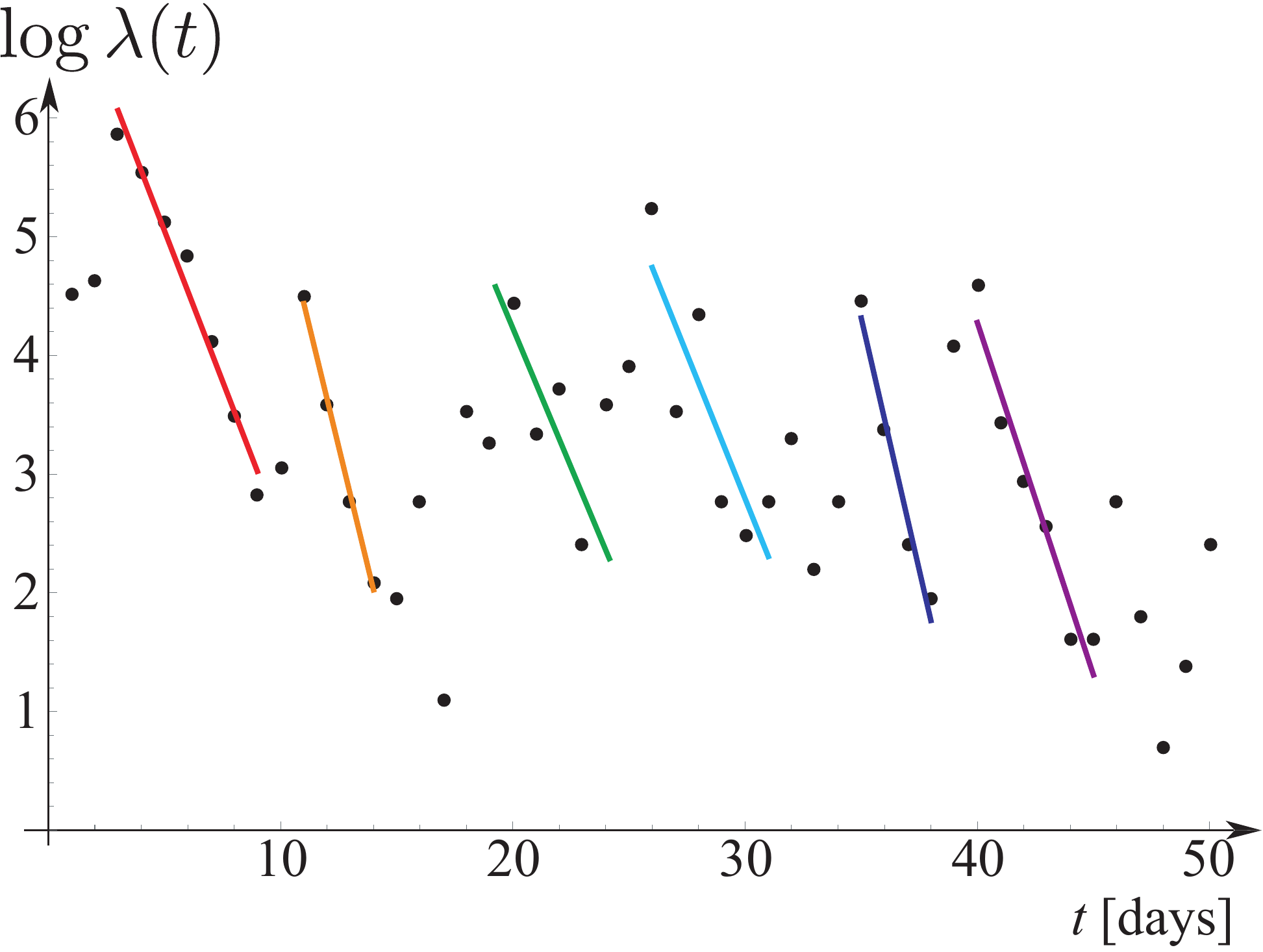} 
 \end{center}
\caption{\label{Fig:DataHRSlopes}  The semi-log plot shows the exponential decay after reaching peak activity. The lines correspond to the fit: $\log \lambda(t) = -\omega t + b,$ showing the desired exponential decay of the model. The slopes are summarized in Table \ref{Table} in the column $\omega^{(i)}$.} \end{figure}

\item  We highlight the asymmetry of the events in time: a sharp lift-off prior to the peak of disorder, followed by a slow exponential decay afterwards, displaying a highly nonlinear behavior. 
\item  A peculiarity of the Chilean unrest movement is its temporal extent. In prior studies\cite{Burbeck1978,bonnasse2018epidemiological}, rioting is characterized by a single huge event that eventually dissipates completely. Yet, as shown in Fig. \ref{Fig:DataHR}, the Chilean case displays a significant number of sizeable events. More importantly, we can observe that they were distributed throughout a month and a half; furthermore, several more small fluctuations subsequently took place during the summer period, up to mid-March 2020. The modelling of these non-predicted events represents an interesting challenge in the context of epidemiological models.
 
\end{enumerate}

\section{Epidemic-like model for riots}
\label{Sec:Epidemic}
Unlike for an infectious disease- which needs close contact between people for it to spread- social media may affect how protests and hence rioting events propagate, since its use can have a positive effect on  participation in political and civic events\cite{boulianne2015social}. This has been observed in particular in the case of some Chilean past massive protest movements\cite{scherman2015student}. It has also been studied that influencing peoples' opinions remains feasible within social media\cite{weeks2017online}.  In particular, Valenzuela  \textit{et al.\@} \cite{valenzuela2016social} study how social media relates to protest participation in Latin America and show that the use of social media for political purposes significantly increases the likelihood of protests, and reduces participation gaps associated with different social groups. Therefore, in the presence of social media communication, a first modelling approach that assumes homogeneous mixing for the susceptible (potential rioters)  and ``infected'' (rioters) population may be reasonable, \textit{i.e.} every individual has the same probability of interacting with every other individual and get ``infected'' with a rioting activity.
\subsection{Model, definitions and basic properties}\label{SubSec:Model}

Almost a hundred years ago, Kermack and McKendrick\cite{Kermack} introduced a mathematical model for epidemics, which gave the base for the model that is known today as the $SIR$ model for describing the dynamics of infectious deseases.  The model is governed by three variables: $I(t)$ represents the number of ``infected'' individuals at time $t$, $S(t)$ is the number of individuals who are susceptible to the disease and could move to the ``infected'' class, and $R(t)=N-S(t)-I(t)$ is the number of removed individuals.  The dynamics of the model is solely determined by $I(t)$ and $S(t)$, since $N$, the total population, is constant. The equations for susceptible and infected individuals of the $SIR$ model therefore read \cite{Kermack} (we adapt in the current paper the notations proposed by Ref. \onlinecite{bonnasse2018epidemiological})
\begin{eqnarray}
\frac{d}{dt} I(t) &=&-\omega I(t) + S(t) P(s\to i,t) \nonumber \\
\frac{d}{dt} S(t)& =-&S(t) P(s\to i,t)\nonumber
\end{eqnarray}
where $\omega$ is a decaying rate of the infected population and $ P(s\to i,t) $ represents the  transmission rate at which a susceptible individual becomes infected. This latter term is assumed to be proportional to the number of individuals currently infected - that is, $ P(s\to i,t) =\kappa I(t)$, with $\kappa$ a parameter with units of 1/time $\times$ 1/(number of individuals).

By analogy with  the above approach, Burbeck \textit{et al.\@} \cite{Burbeck1978} introduced an epidemic-like model for riots, in which $I(t)$ represents the total number of active rioters, and $S(t)$ denotes the number of individuals susceptible to join the disorder. Further, it is assumed that the number of riot events, $\lambda(t)$, is proportional to the number of rioters, {\it i.e.} (the symbol $ := $ indicates a definition)
$$\lambda(t) := \alpha I(t),$$
where $\alpha$ is assumed to be constant with units of (number of events)/(number of individuals). Equivalently, the potential `supply' of riot events, $\sigma(t)$, is defined in terms of the volume of available individuals:
$$\sigma(t) := \alpha S (t).$$
Then, the model proposes: 
\begin{eqnarray}
\frac{d \lambda}{d t} &=&- \omega \lambda + \beta \sigma \lambda , \label{Eq:lambda} \\
\frac{d \sigma}{d t} & = & -\beta \sigma \lambda .\label{Eq:sigma}
\end{eqnarray}
Here, $\omega$ and $ \beta=\frac{\kappa}{\alpha}$, with units of 1/time and  1/time $ \times $ 1/events respectively, are parameters of the problem, representing the exit rate from the riot events class and the transmission rate per riot event. We emphasize that the parameters $\alpha$ and $\kappa$  are unknown in this context. Finally, this set of ordinary differential equations (o.d.e.) is complemented by the initial conditions:
\begin{eqnarray} \sigma(t_0)=\sigma_0 \quad \& \quad \lambda(t_0) = \lambda_0.\label{Eq:Initial}\end{eqnarray}

\begin{figure}[h]
%(a)  \includegraphics[width=7cm]{Integral-eps-converted-to.pdf} \vskip 0.5 cm (b)  \includegraphics[width=7cm]{yvst-eps-converted-to.pdf}
(a)  \includegraphics[width=3.5cm]{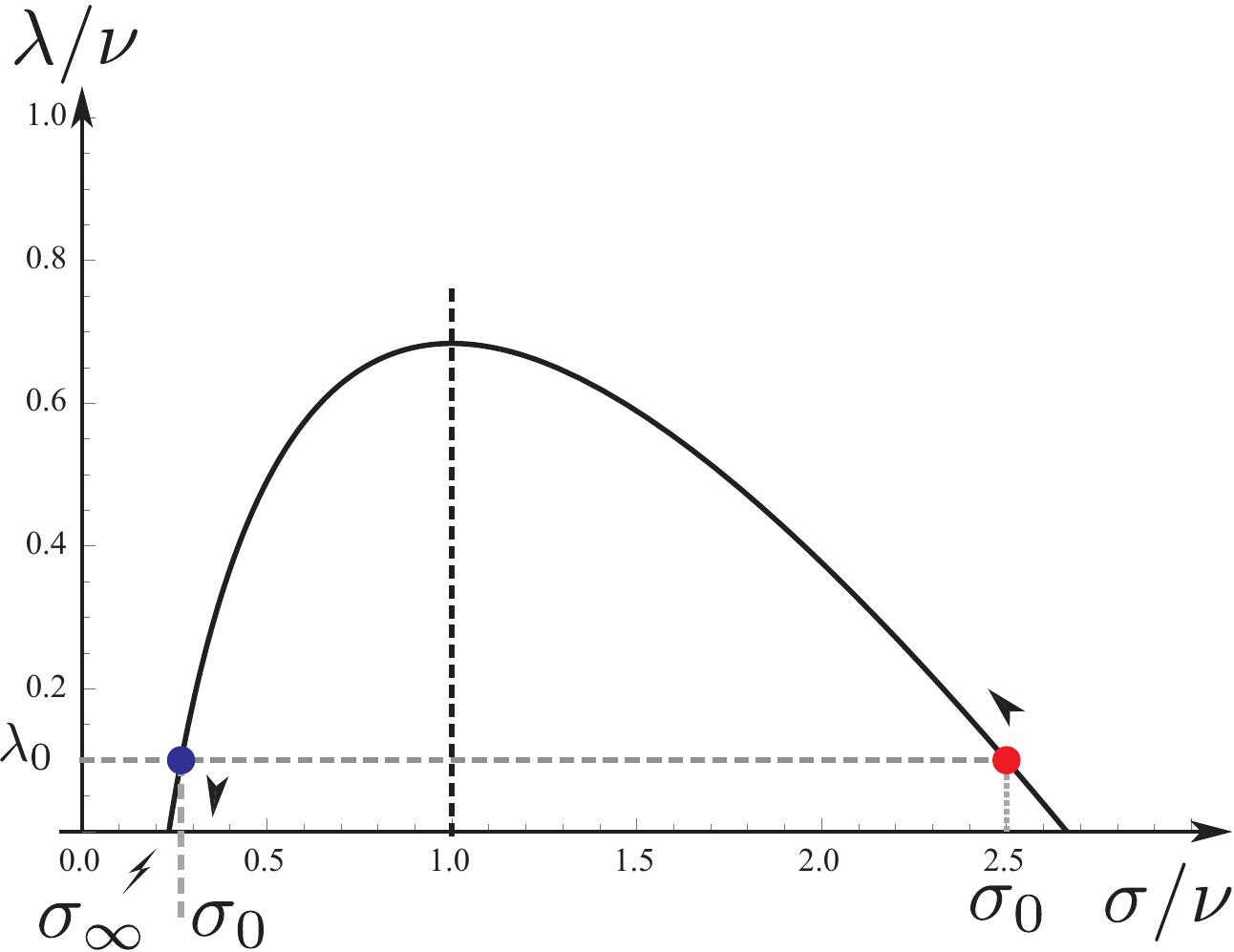}  \, (b)  \includegraphics[width=3.5cm]{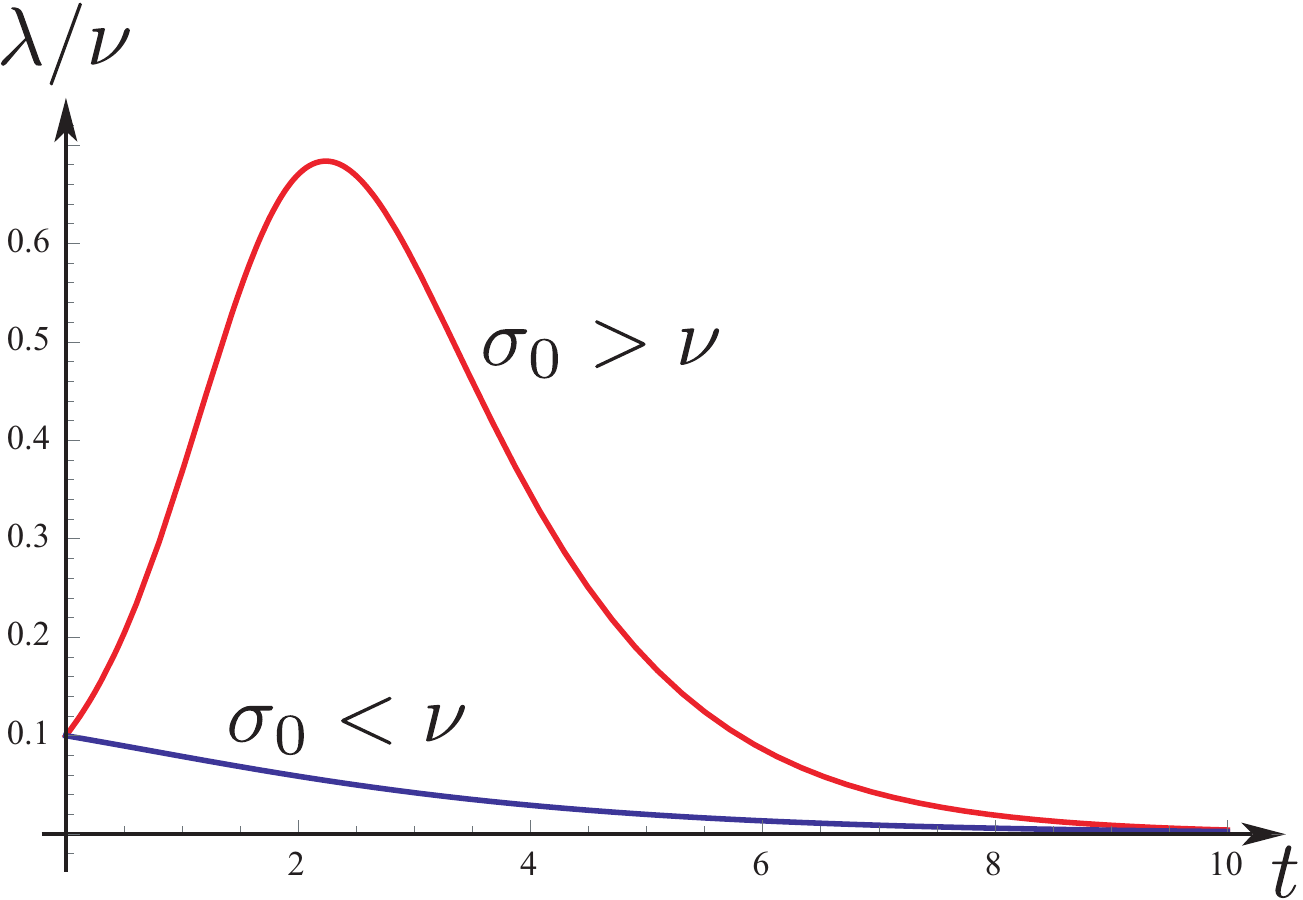}
\caption{\label{Fig:Integral}  (a) Plot of the conserved quantity (\ref{Eq:Integral}). The maximum of (\ref{Eq:Integral}) is  for $\sigma/\nu=1$,  and the flow of the dynamics is from right to left as indicated by the arrows. The flow shows that as $t\to \infty$, $\lambda(t)\to 0$ and $\sigma(t)\to \sigma_\infty$.   If $\mathcal{R}_0=\sigma_0/\nu > 1$, then, the trajectory in the $(\lambda,\sigma)$ plane goes from the starting point to the left reaching $\lambda=0$. While, if $\mathcal{R}_0=\sigma_0/\nu < 1$ the trajectories end directly at $\lambda=0$ without passing by the maximum of the curve. (b) Temporal evolution of the number of riot events for two different initial conditions, corresponding to the red and blue points of (a) and fixed $\nu$ value. }
\end{figure}

As a first consequence of the model, one has that the number of inactive individuals, which are the ones susceptible to join a riot, decreases strictly in time, therefore the asymptotic dynamics is: $\sigma(t) \to \sigma_\infty$, thus, $ \lambda(t) \to 0$ as $t\to \infty$ as well. More importantly, this system is integrable; this property solves exactly the problem, and provides a simple frame of analysis for the contagion model. Dividing (\ref{Eq:lambda}) by (\ref{Eq:sigma}) gives
$
\frac{d \lambda}{d \sigma}=    \frac{\omega}{\beta}  \frac{1}{\sigma} -1  , 
$
and thus 
\begin{eqnarray}
\lambda+ \sigma - \nu  \log \sigma = C .  \label{Eq:Integral0} 
\end{eqnarray}
Here we define the shorthand notation $\nu =  \frac{\omega}{\beta} $.
The constant $C$ should be fixed by the initial conditions (\ref{Eq:Initial}), so that:
\begin{eqnarray}
\lambda &=&\lambda_0+  \nu \log\left(\frac{ \sigma}{\sigma_0} \right)  - (\sigma -\sigma_0)   .  \label{Eq:Integral} 
\end{eqnarray}
 This integral (\ref{Eq:Integral}) is sketched in Fig. \ref{Fig:Integral}-(a).

In principle the integral (\ref{Eq:Integral})  solves the problem. Nevertheless, to find the time dependent behavior one replaces (\ref{Eq:Integral}) into  (\ref{Eq:sigma}), thus
$$\frac{d \sigma}{d t} = -\beta \sigma \left(  \lambda_0+\nu \log\left(\frac{ \sigma}{\sigma_0} \right)  - (\sigma -\sigma_0)   \right) ,$$
that is 
%\begin{eqnarray}\Phi(\sigma;\sigma_0,\lambda_0; \nu)&=&\int_{\sigma_0}^\sigma \frac{d\sigma}{\sigma \left(  \lambda_0+ \nu\log\left(\frac{ \sigma}{\sigma_0} \right)  - (\sigma -\sigma_0)   \right)} \nonumber\\
%&=& - \beta (t-t_0) .\label{Eq:Integral2} 
%\end{eqnarray}
 \begin{widetext}
\begin{eqnarray}\Phi(\sigma;\sigma_0,\lambda_0; \nu)&=&\int_{\sigma_0}^\sigma \frac{d\sigma}{\sigma \left(  \lambda_0+ \nu\log\left(\frac{ \sigma}{\sigma_0} \right)  - (\sigma -\sigma_0)   \right)} = - \beta (t-t_0) .\label{Eq:Integral2} 
\end{eqnarray}
 \end{widetext}
Finally, one computes $\lambda(t)$ through Eq.  (\ref{Eq:Integral}).

\subsection{Activation of rioting epidemics and their prevention}\label{SubSecc:Activation}
It is interesting to answer the question about what happens if we introduce a small number of rioters ($\lambda_0/\alpha$) into a population of susceptibles ($\sigma_0/\alpha$); will there be an epidemic of riot events? 
The dynamics depends crucially on the initial value $\sigma_0$; more precisely, they depend on whether $\sigma_0$ is greater or smaller than $\nu=\omega/\beta$. Indeed, $d\sigma/dt<0$ for all $t$ (since the right-hand side of eqn. (\ref{Eq:sigma}) is negative) and $d\lambda/dt>0$ if and only if $\sigma>\nu$ (see the right-hand side of eqn. (\ref{Eq:lambda})). Therefore, $\lambda(t)$ increases as long as $\sigma>\nu$, but since $\sigma(t)$ is decreasing, $\lambda(t)$ eventually will decrease and approach zero. Hence, if $\sigma_0<\nu$, then  $\lambda(t)$ decreases strictly to zero exponentially for all $t$ (the right-hand side of eqn. (\ref{Eq:lambda}) is always negative), which means that there is no epidemic of riot events. On the other hand, if $\sigma_0>\nu$, then the right-hand side of eqn. (\ref{Eq:lambda}) is first positive and hence $\lambda(t)$ first increases to a maximum, which is attained when $\sigma=\nu$, and then the right-hand side of eqn. (\ref{Eq:lambda}) becomes negative so that $\lambda(t)$ decreases strictly to zero exponentially. These latter dynamics correspond to the occurrence of an epidemic of riot events. In this case, the initial number of susceptible individuals (potential rioters) was sufficient for active rioters to ``infect'' enough susceptible individuals to join the riots and produce riot events.  That way, riot events invade the population up to a certain point (maximum), and then eventually decay.
These simple dynamics are represented in Fig. \ref{Fig:Integral} (b), for $\sigma_0<\nu$ (no epidemic) and $\sigma_0>\nu$ (epidemic). 

Regardless of the initial value of $\sigma$, $\lambda$ decreases to zero as 
\begin{eqnarray}\lambda(t)  %  \approx \lambda_0+ \nu\log\left(\frac{ \sigma(t)}{\sigma_0} \right)+ \sigma_0
\approx   \lambda_*  e^{-\omega (t-t_*)} \quad t\to\infty.\label{Eq:AsymptoticSolution} 
\end{eqnarray}
Therefore, the long time behavior of $\lambda $ is to decay exponentially in time, with decaying rate simply given by the parameter $\omega$. For this reason, $\omega$ may be seen as the inverse of the mean lifetime of the riot event, and we thus explicitly define this lifetime, $\tau$, as $\tau = 1/\omega.$

The quantity $\sigma_0/\nu=\beta \sigma_0 \times \frac{1}{\omega} = \beta \sigma_0\tau $ is a threshold quantity, which in epidemiological modelling is called the \textit{Basic Reproduction Number} and is usually denoted by $\mathcal{R}_0$ (See Refs. \onlinecite{Heesterbeek2002,Brauer2012}). Its value determines the dynamics of the system; since  a riot epidemic occurs when $\sigma_0/\nu>1$, which is equivalent to $\mathcal{R}_0>1$, and no riot epidemic happens when $\mathcal{R}_0<1$. The interpretation of $\mathcal{R}_0=\beta \sigma_0 \tau$ in the context of riot activity is that it is the average number of secondary riot events caused by a single riot event introduced into a population of susceptible individuals (potential rioters) of size $\sigma_0/\alpha$ over the duration of that single riot event ($\tau = 1/\omega$). This can be explained step by step in the following way: $\beta$ is the riot events transmission rate, which is a parameter that is equal to the product of the contact rate of an active rioter with susceptible rioters times the probability per contact that such a contact produces ``infection'', {\it i.e.} produces a new active rioter that generates rioting events. Hence, one riot event produces $\beta \sigma_0$ new riot events per unit of time in a population with $\sigma_0/\alpha$ susceptible individuals ($\sigma_0$ potential riot events). Since $\omega$ is the parameter that represents the exit rate from the riot events class, $\tau=1/\omega$ is the mean lifetime of each riot event (see eqn. (\ref{Eq:AsymptoticSolution}) and its explanation). Therefore, $\beta \sigma_0 \tau$ is the average number of new riot events in a population with $\sigma_0/\alpha$  susceptible individuals ($\sigma_0$ potential riot events) produced by one riot event during the time it lasts. 

In this simple model, the transition from a dynamic that displays an epidemic of riot events to one that does not, is simply characterized by the initial value of potential riot events ($\sigma_0$), the riot transmission rate ($\beta$) and the duration of a riot event ($\tau$). In particular, if $\beta$ and $\tau$ are fixed, the bifurcation scheme of the system is characterized by a change in the initial state of potential rioters.

Models like the ones presented in this section can give qualitative insights on rioting dynamics that could help policymakers to make informed decisions. As discussed above, the threshold quantity $\mathcal{R}_0=\beta \sigma_0 \tau$ determines riot event dynamics. In order to prevent riot events epidemics in a population with a fixed number of $\sigma_0/\alpha$ potential rioters, a public safety policy would need to implement measures that reduce $\tau$ and/or $\beta$ such that $\mathcal{R}_0<1$, in which case a riot events epidemic would be prevented. 

 \begin{figure}[h]
 (a)  \includegraphics[width=3.5cm]{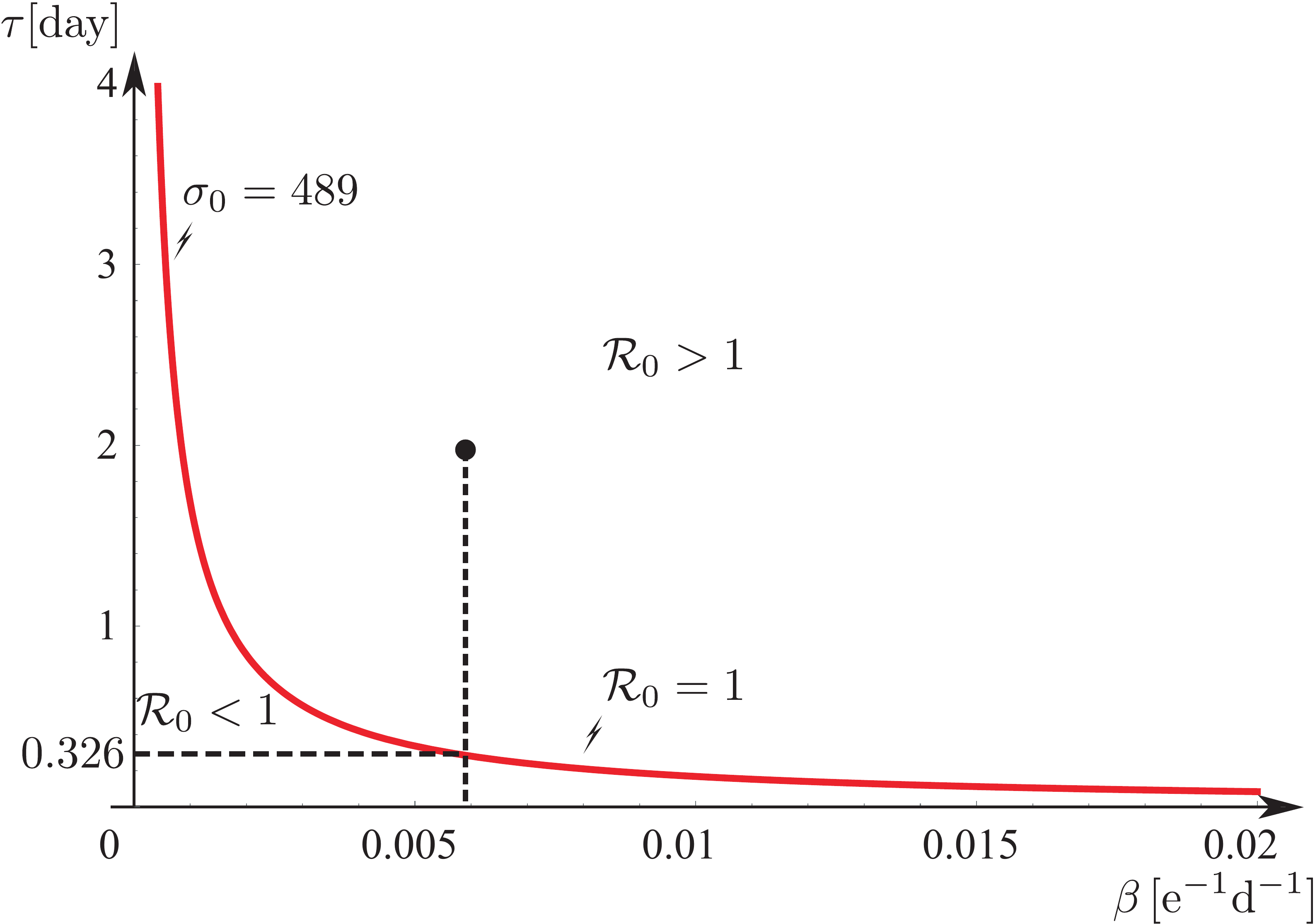}  \, (b)  \includegraphics[width=3.5cm]{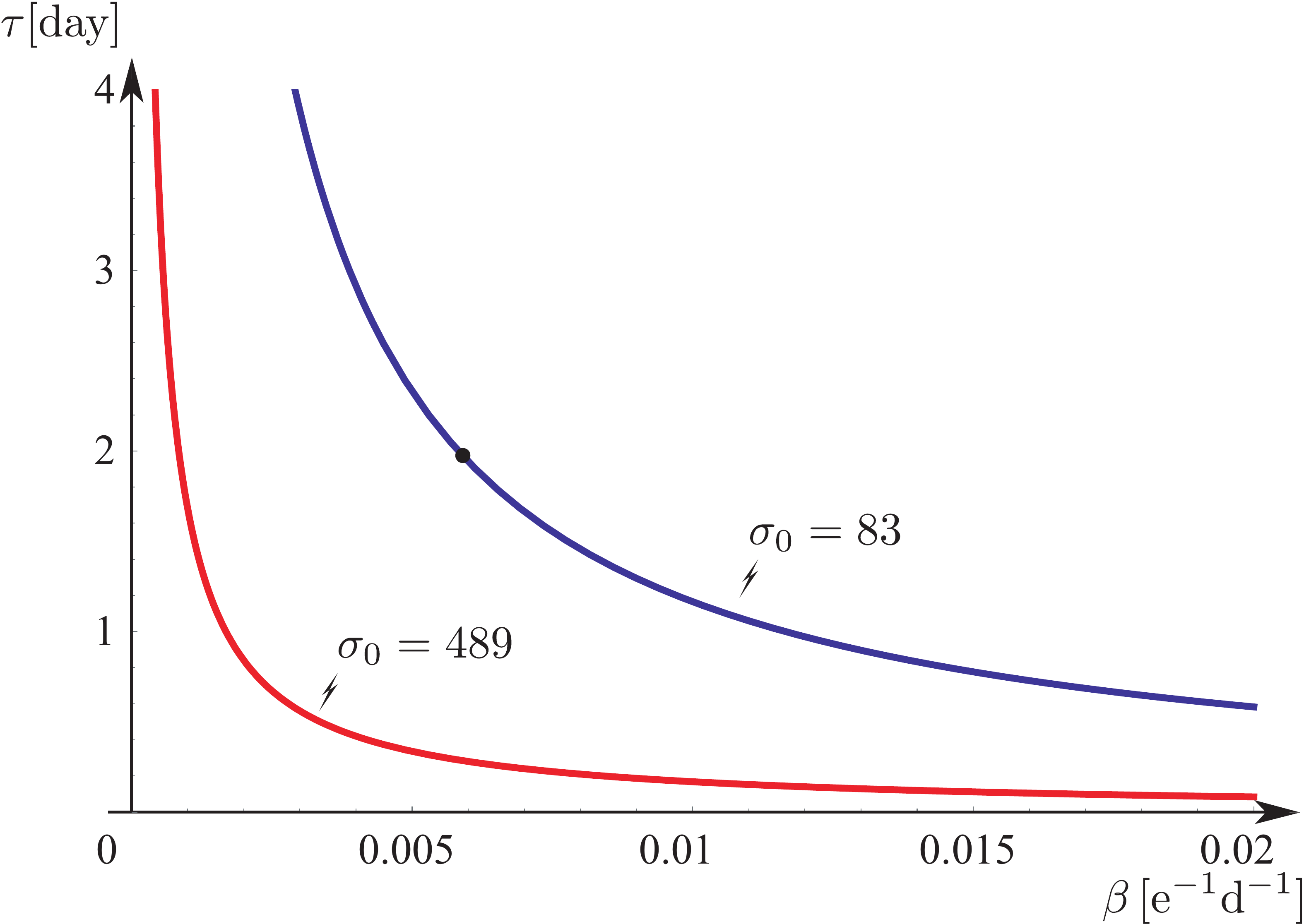}
 % (a) \includegraphics[width=7cm]{Discussion2-eps-converted-to.pdf}\\ \vskip 0.5 cm(b) \includegraphics[width=7cm]{Discussion1-eps-converted-to.pdf}
\caption{\label{Fig:Disc1}   (a) Plot of  the threshold curve $\mathcal{R}_0=\beta \sigma_0 \tau =1$ in the $(\beta,\tau)$ plane for riot event epidemics. The red curve represents the threshold curve for $\sigma_0 = 489$ [e]  that divides the two regions where $\mathcal{R}_0>1$ and $\mathcal{R}_0<1$. The dot represents the case of the major epidemic event (1) of Fig. \ref{Fig:DataHR}, where the fitting parameters are   $\omega= 0.521 $ [1/d], $\beta = 0.0063$  [$\rm e^{-1} d^{-1}$], and $\sigma_0 = 489$ [e]  from Table \ref{Table}  which produce $\mathcal{R}_0=5.882>1$.
(b) In addition of (a) it is show blue curve for different $\sigma_0 =83.2 $ [e] which produce $\mathcal{R}_0=1$.  }
  \end{figure}

Fig. \ref{Fig:Disc1}(a) shows the threshold curve $\mathcal{R}_0=1$ for $\sigma_0=489$ [e] that corresponds to the first peak in Fig. \ref{Fig:DataHR}.  As pictured, the curve divides the parameter space $(\beta, \tau)$ into two regions: the region above the curve represents parameter sets for which riot event epidemics occur ($\mathcal{R}_0>1$) and the region below the curve represents parameter sets for which no epidemics occur ($\mathcal{R}_0<1$).

The major riot events epidemic (the first peak in Fig. \ref{Fig:DataHR}) corresponds to a parameter set of a mean lifetime duration per riot event of $\tau\approx 1.33$ [d], a transmission rate $\beta = 0.0063$  [$\rm e^{-1} d^{-1}$] per riot event and a population of susceptible rioters that would be responsible for $\sigma_0=489$ [e] potential riot events (see Table \ref{Table}). This riot events epidemic has a basic reproduction number of $\mathcal{R}_0=5.88$ (see Table \ref{Table}), which is indeed greater than 1 (epidemic) and its parameter set is represented by the dot in Fig. \ref{Fig:Disc1}(a). In order to prevent future riot events epidemics of this form, in a population with $\sigma_0=489$ [e] potential riot events, a public safety policy that sufficiently reduces the mean lifetime duration of a riot event ($\tau$) would be able to prevent such an epidemic. Fig. \ref{Fig:Disc1}(a) shows that reducing $\tau$ below $0.326$ [d], while fixing $\beta$, produces a rioting situation in the region where $\mathcal{R}_0<1$ and hence a riot event is prevented. 

 Some strategies to prevent outbreaks of contagious rioting involve measures that target the transmission rate $\beta$. This could be achieved by reducing the contact rate between active rioters and the susceptible population, or by lowering the probability that an active rioter, upon contact with a susceptible individual, ``infects'' him or her, creating more rioters and hence more riot events. The effectiveness with which ideas can be spread via social media and the intrinsic characteristics of how a population communicates (physical or virtual) may be key to predict the behavior of an upcoming riot activity. 

If the number of potential riot events existing initially in a population ($\sigma_0$) changes, the public safety policy that reduces the duration of a riot event or targets the contact rate between individuals could change. Fig. \ref{Fig:Disc1}(b) shows that in a population where now $\sigma_0=83$ [e], the threshold curve changes and hence the threshold values for the parameters, which are needed to pass from the region $\mathcal{R}_0>1$ to the region $\mathcal{R}_0<1$, change.

\subsection{Hamiltonian dynamics}
The existence of the conservation quantity (\ref{Eq:Integral0}) indicates that the dynamics follows  a tangential flow from the conserved quantity. In other words, the system may be written in terms of a Hamiltonian dynamics.
Let be, $$H= \beta( \lambda + \sigma ) -  \omega  \log \sigma ,  $$
 then equations (\ref{Eq:lambda}) and (\ref{Eq:sigma}) may be written as:
 \begin{eqnarray}
\frac{1}{\lambda} \frac{d \lambda}{d t} &=& \sigma \frac{\partial H}{\partial \sigma}, \label{Eq:lambda2} \\
\frac{1}{ \sigma} \frac{d \sigma}{d t} & = & -\lambda \frac{\partial H}{\partial \lambda}. \label{Eq:sigma2}
\end{eqnarray} 
Thus, taking the canonical variables $(x,y)$ as:
 \begin{eqnarray}
 x &=& \log \lambda  , \label{Eq:DefX} \\
y &=& \log \sigma  , \label{Eq:DefY}
\end{eqnarray} 
 and a Hamiltonian 
 \begin{eqnarray}
H &=& \beta\left( e^{x} + e^y\right)  - \omega  y, \label{Eq:DefH}
\end{eqnarray} then,
one writes the Burbeck, \textit{et al.\@} epidemic model with a Hamiltonian structure:
 \begin{eqnarray}
\frac{d x}{d t} &=&  \frac{\partial H}{\partial y} = -\omega +  \beta e^y  , \label{Eq:HamiltonX0} \\
\frac{d y}{d t} &=&  -\frac{\partial H}{\partial x}= - \beta e^x  . \label{Eq:HamiltonY0}
\end{eqnarray} 
Moreover, as already shown after (\ref{Eq:Integral}), the model is integrable. This fact will be pertinent in the following applications.

\subsection{Parametric periodic and stochastic forced variations of Burbeck,  \textit{et al.\@} epidemic model.}\label{SubSec:forcing}

By construction, the model equations (\ref{Eq:lambda}) and (\ref{Eq:sigma}) cannot predict more than one riot event. Indeed, because the number of potential rioters, $\sigma$, decreases strictly in time (see eqn.  (\ref{Eq:sigma})), the graphical scheme presented in Fig. \ref{Fig:Integral}  shows that the dynamics does not present a returning point. Therefore, there is no way to revert the trajectory in the $(\lambda,  \sigma)$ phase portrait. We conclude that, to reach a dynamical behavior with various events, one needs to modify the original model by including other basic phenomena. Essentially, the number of potential rioters must increase in some way, which can be done using different source terms. The simplest way is to introduce a linear growth rate of the individuals that eventually may join a riot, $\sigma$; more precisely, a $\gamma\sigma $ term on the right-hand side of  (\ref{Eq:sigma}). This model is equivalent to the well known Lotka-Volterra population dynamics model. Another possibility is a time dependent forcing. We summarize these different possibilities in the following short hand notation:

\begin{eqnarray}
\frac{d \sigma}{d t} & = & -\beta \sigma \lambda + \sigma(t)  \left\{ \begin{array}{c}
 \gamma \\
f(t) \\
\sqrt{\eta} \,  \xi(t) 
\end{array}
 \right. .\label{Eq:sigmaForced}
\end{eqnarray}
%for instance: $f(t) = \frac{ A}{2}  \left(1-\cos(\Omega t)\right)$
Here $f(t)$ denotes a periodic forcing, that we will define precisely later on, and $\xi(t)$ is a stochastic variable which we consider as a $\delta$-correlated white noise. More formally, the first two moments are $\left< \xi(t)\right>=0$, and $\left< \xi(t) \xi(t')\right>=\delta(t-t')$. Here  the function $ \delta(t)$ denotes a Dirac $\delta$-function which simulates extremely short-time correlation. More specifically, $ \delta(t)$ is zero at every time except at $t=0$, and its integral satisfies $$ \int_{-\infty}^{t} \delta(s) \, ds =\left\{ \begin{array}{cc}
 0 &\quad t< 0  \\
1 &\quad t>0
\end{array}
 \right.  
 .
 $$

Therefore, equations (\ref{Eq:lambda}) and (\ref{Eq:sigmaForced}) may be written in the following form after a suitable change of variables (\ref{Eq:DefX}) and (\ref{Eq:DefY}) to a set of Hamilton equations plus an additive forcing:
\begin{eqnarray}
\frac{d x}{d t} &=&  \frac{\partial H}{\partial y}  , \label{Eq:HamiltonX} \\
\frac{d y}{d t} &=&  -\frac{\partial H}{\partial x} +\left\{ \begin{array}{c}
 \gamma \\
f(t)\\
\sqrt{\eta} \,  \xi(t) 
\end{array}
 \right.   , \label{Eq:HamiltonY}
\end{eqnarray} 
where 
$ H $ is given by (\ref{Eq:DefH}).

\subsubsection{The case of constant forcing.}
In this case the epidemic model becomes a special Lotka-Volterra system that displays a pure oscillatory behavior. The dynamics are periodic and quite trivial; more importantly, they are un-realistic in the current context and so we will not discuss this case further.

\subsubsection{The case of periodic forcing.}
As stressed, the social movement possesses a natural weekly period. A natural way to model that is by taking the periodic excitation:
 \begin{eqnarray}
f(t) &=& f_0 \sum_{l=1}^\infty \delta(t-l T) . \label{Eq:PeriodicForcing}
\end{eqnarray} 
Here $f_0$ is the forcing intensity,  $T$ is the period, and the $\delta$-function emulates a time localized excitation of the variable $\sigma$; a kind of periodic kick. The equations of motion for this {\it kick-epidemic model} read
 
  \begin{eqnarray}
\frac{d x}{d t} &=& -\omega +  \beta e^y  , \label{Eq:KickX} \\
\frac{d y}{d t} &=&   - \beta e^x +  f_0 \sum_{l=1}^\infty \delta(t-l T)  . \label{Eq:KickY}
\end{eqnarray} 

 The whole solution $(x(t),y(t))$ for all $t$ is obtained by imposing at each
$t=ld$ the continuity of the function $x(t)$ and a jump condition on $y(t)$ which is obtained directly after an integration of Eq. (\ref{Eq:KickY}) in a small
interval, $t\in (lT^{-},lT^{+})$, around 
$t=l T $. 
The jump conditions read
\begin{eqnarray}
x(lT^{+}) &=& x(lT^{-}) \nonumber \\
y (lT^{+})&=&y (lT^{-}) + f_0, \label{salto}
\end{eqnarray} 
where $y(lT^{+})$ is the value of $y(t)$
just after  $t=lT$ and
$y(lT^{-})$ is the value of $y(t)$ just before
 $t=lT$.
In the original variables, 
if $f_0$ is positive, then the kick will increase the number of potential rioters, thus making it possible to re-start a riot if $\sigma (lT^{+})=e^{f_0} \sigma (lT^{-})  > \nu $.  On the other hand,  the number of events, {\it i.e.} the $\lambda$ variable, is a continuous function at the kick.

\begin{figure}
(a)  \includegraphics[width=7cm]{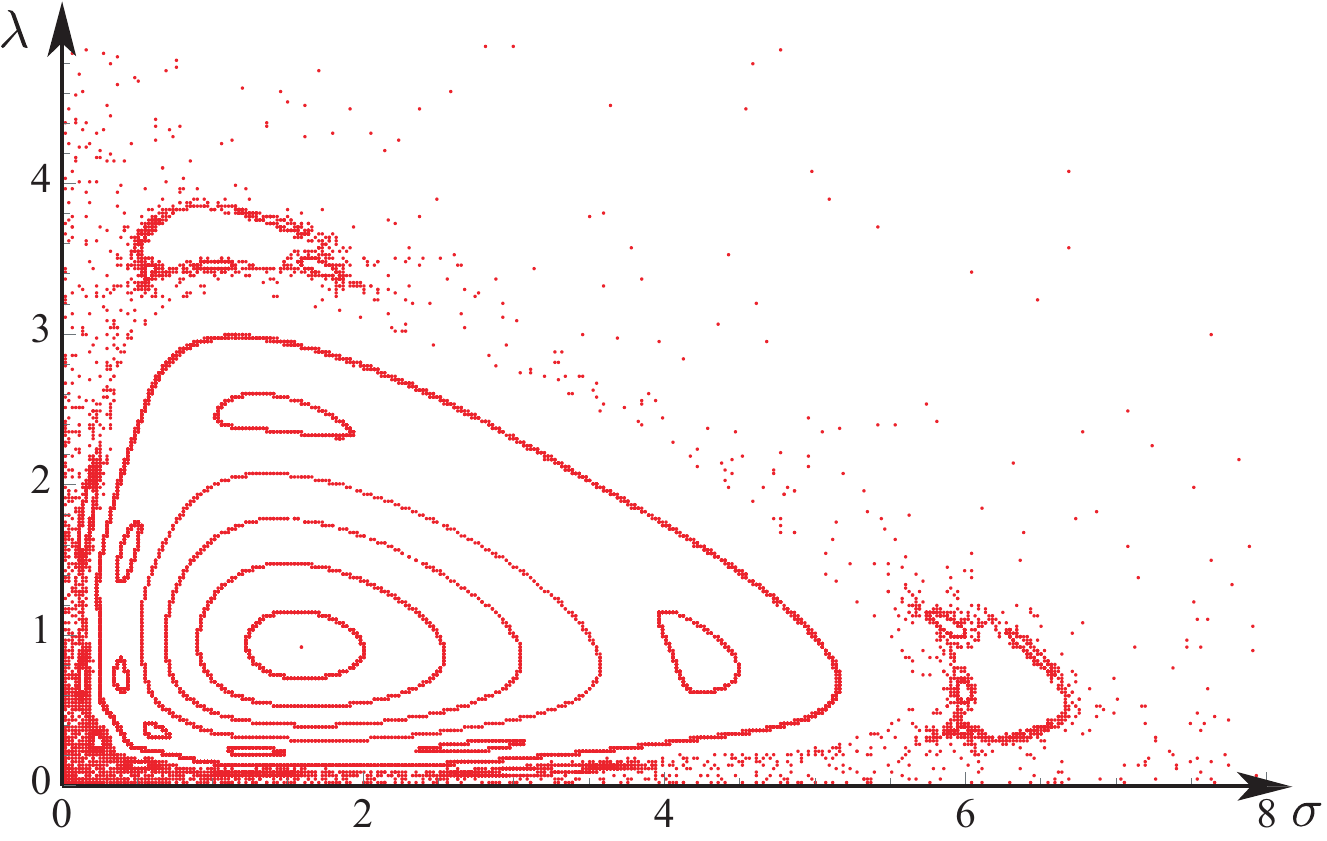} \\ (b)  \includegraphics[width=7cm]{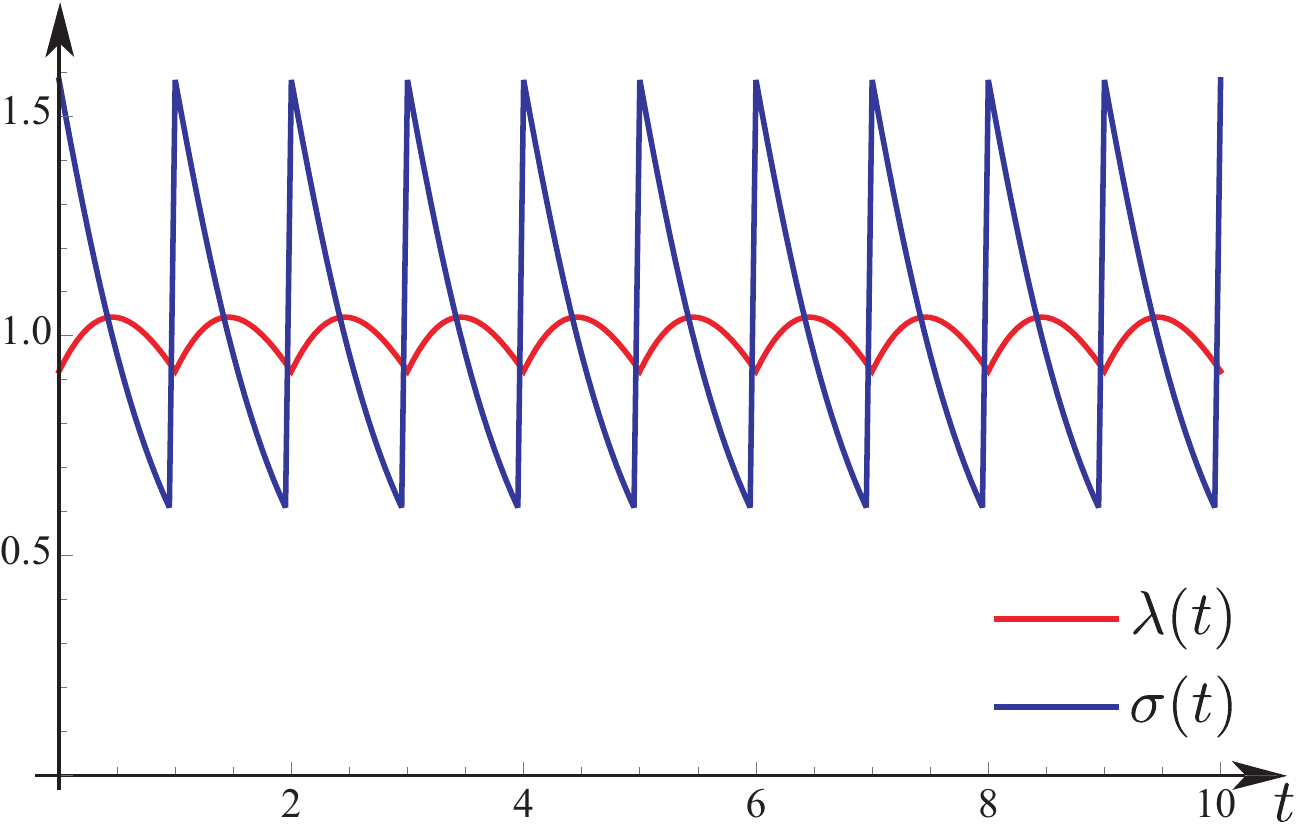}\\
(c)  \includegraphics[width=7cm]{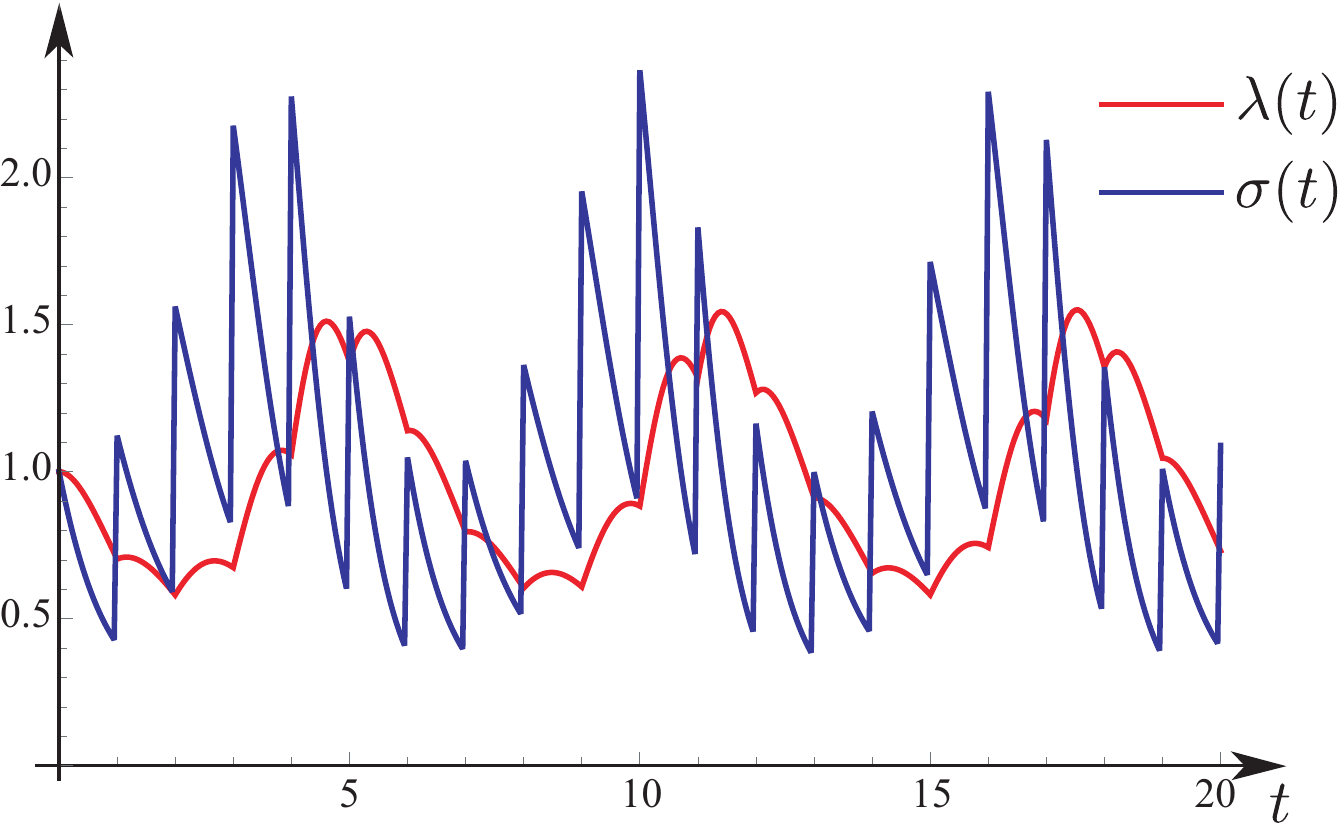} \\ (d)  \includegraphics[width=7cm]{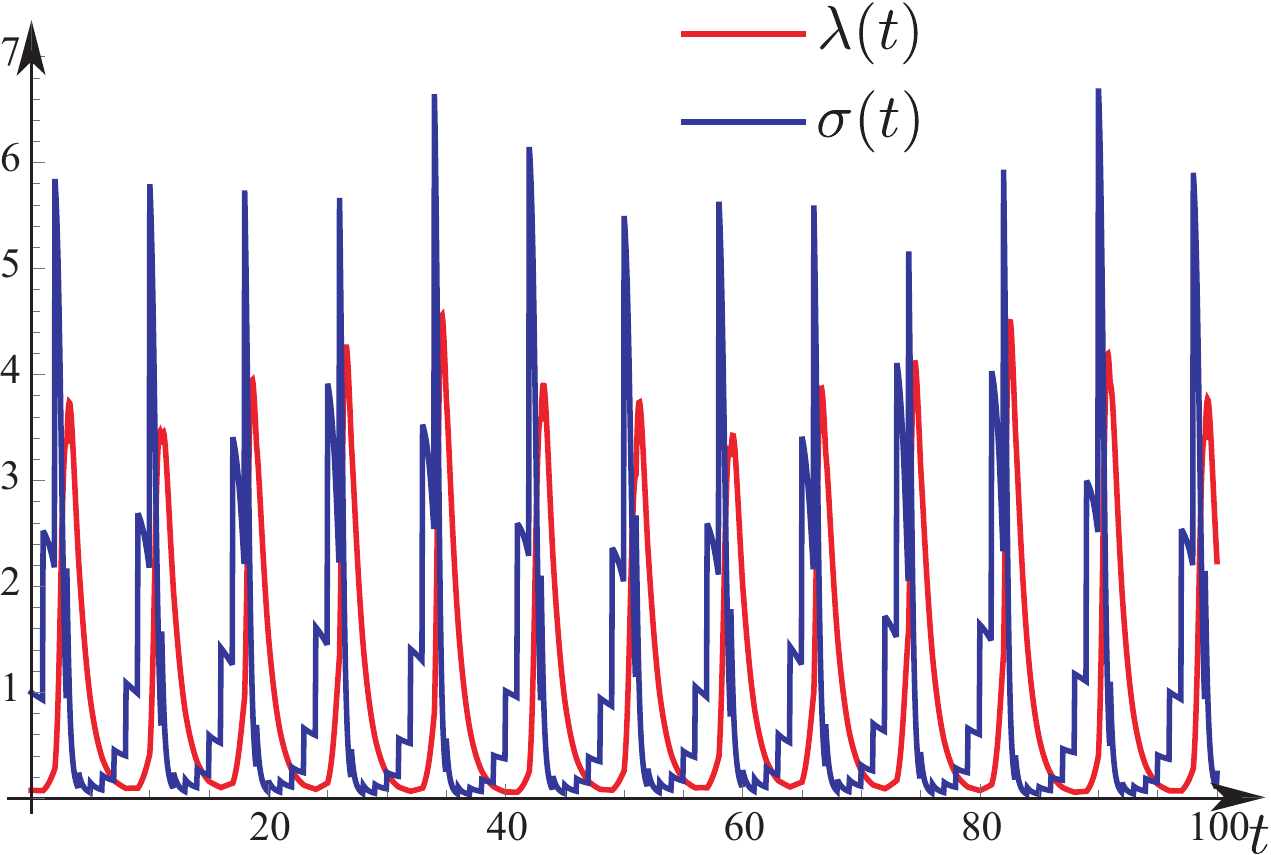}
\caption{\label{Fig:PoincareMap} (a) Poincare map for the couple  $(\lambda(lT^+),\sigma(lT^{+}))$, for the parameters: $\omega=\beta$, $T=1$ and $f_0=1$  and for different values of the initial conditions: $(\lambda(0),\sigma(0))$. It is seen that the system displays a periodic orbit, which corresponds to a fixed point in the map (See (b)). Around the fixed point one notices the existence of  closed orbits that represent a quasi-periodic behavior. Finally,  because of the external orbits break-up into a domain of chaotic behavior. The next plots show the temporal behavior of the variables $\lambda(t)$ and $\sigma(t)$ for various initial conditions such that the dynamics displays: (b) a periodic behavior for $(\lambda(0),\sigma(0))\approx ( 0.9194125, 1.58198)$; (c) a quasi-periodic behavior $(\lambda(0),\sigma(0)) \approx (1, 1)$; and (d) a chaotic behavior $(\lambda(0),\sigma(0))\approx (0.075, 1)$.  }
\end{figure}

From a mathematical perspective,
when $f_0 \neq 0$ the system is perturbed by periodic kicks which breaks the integrability of the system.
The value of $H$, defined in Eq. (\ref{Eq:DefH}), changes at each kick but takes a
constant value $H_{l}$ over each interval $t \in (l T ,(l+1)T)$ where the system is
integrable. The equations
(\ref{Eq:KickX}) and (\ref{Eq:KickY})  can be solved in each of these intervals, with the formal solution being given by (\ref{Eq:Integral}) and (\ref{Eq:Integral2}).

 Using this  formal solution in the $(x,y)$ variables, together with the jump condition (\ref{salto}), we can construct a
stroboscopic or Poincar\'e map ${\mathcal M}$:
\begin{equation}
%[\lambda((l+1)T^+),\sigma((l+1)T^{+})]={\cal M}[\lambda(lT^+),\sigma(lT^{+})].
(x((l+1)T^+),y((l+1)T^{+}))={\mathcal M}\left(x(lT^+),y(lT^{+})\right),
\label{map} 
\end{equation} 
that provides a mapping from one kick to the next one.

The nonlinear map, ${\mathcal M}$, can in principle be expressed explicitly; however, this is not required for our purposes.
Due to the Hamiltonian nature of the evolution, this map has the symplectic property. Maps with this property have been studied vastly in the past. As we know from the properties of the map (\ref{map}), the solutions are either {\it periodic, quasi-periodic} or {\it chaotic} \cite{ArnoldAvez}.

A {\it periodic} behavior appears as a fixed point in the $(\lambda(lT),\sigma(lT))$ phase portrait of the nonlinear mapping (See Fig. \ref{Fig:PoincareMap}(a)-(b)).
%Of course the trivial solution $R=0$ $R_y=0$, is also a {\it periodic} orbit.
As shown in Fig. \ref{Fig:PoincareMap}(a), it appears that this fixed point is elliptic so that the orbits around it represent  a {\it quasi-periodic} behavior, as may be seen in  Fig. \ref{Fig:PoincareMap}(c). Finally, around the border of the orbits there are {\it chaotic} trajectories as can be seen in Fig. \ref{Fig:PoincareMap}(d).

\subsubsection{The case of a stochastic forcing.}
\begin{figure}[h!]
(a)  \includegraphics[width=7cm]{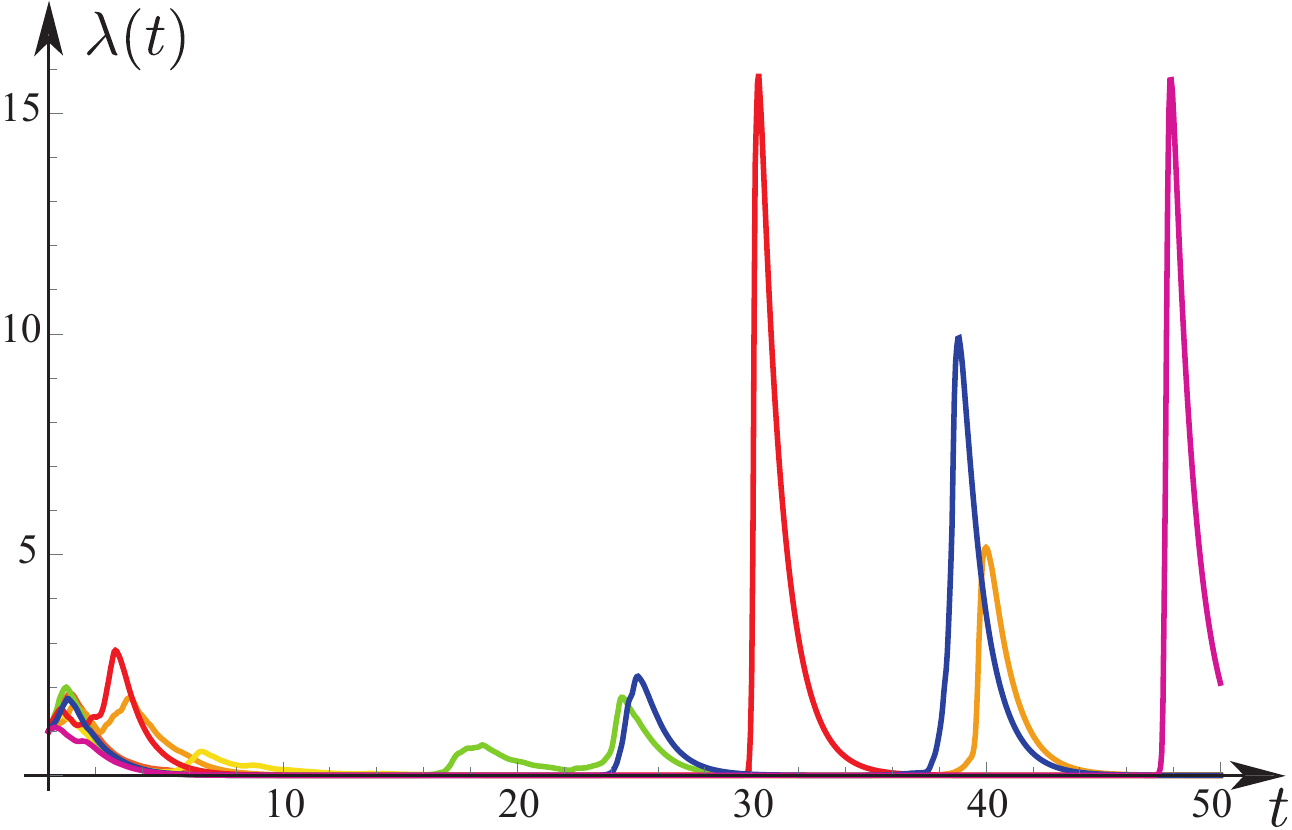} \\ (b)  \includegraphics[width=7cm]{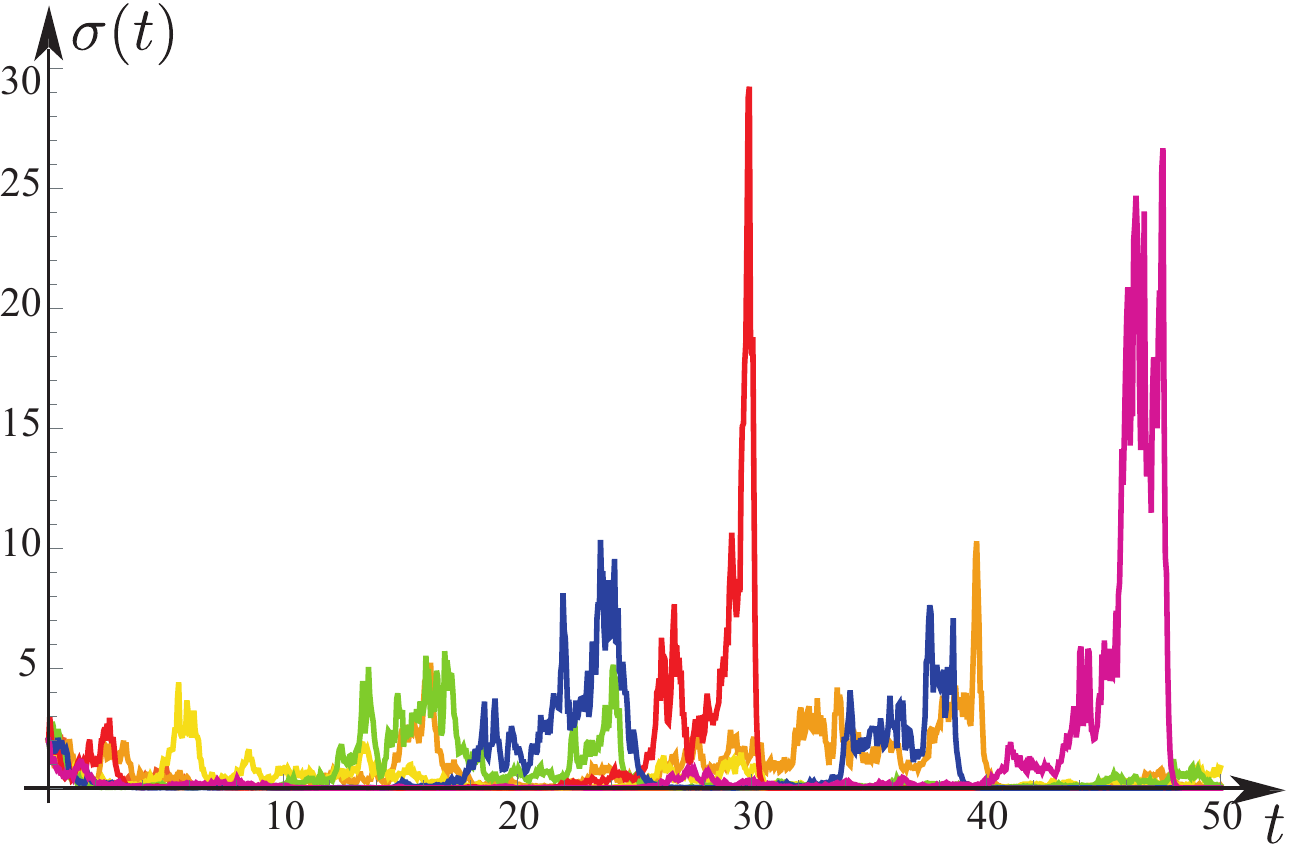}
\caption{\label{Fig:noise}  Numerical simulation of the system (\ref{Eq:lambda}) and (\ref{Eq:sigmaForced}) where  $\xi(t)$ is a white noise. (a) Plot of the temporal evolution of the riot activity, $\lambda(t)$. (b)  Plot of the potential number of rioters, $\sigma(t)$ for the same realizations as in (a). The  parameters are $\omega=\beta=1$ and $\eta=1$. The initial contidion is the same for all realizations: $(\lambda(0),\sigma(0))\approx (1, 2)$. For the numerics we use the  Euler-Mayurama  scheme \cite{EulerMayurama}. }
\end{figure}

In the case of a stochastic forcing, the trajectories are erratic and quite different for various realizations. 
This can be seen in Fig. \ref{Fig:noise}, where seven different stochastic realizations of equations (\ref{Eq:lambda}) and (\ref{Eq:sigmaForced}) (with the stochastic forcing $\xi(t)$) are plotted together.

As is well known, only a probabilistic description makes sense in this kind of differential equations. In the case of eqns. (\ref{Eq:HamiltonX}) and (\ref{Eq:HamiltonY}) (with the stochastic forcing $\xi(t)$), the stationary probability is the so-called Gibbs measure:

\begin{eqnarray}
p_{st} &\sim& e^{-\frac{1}{\eta} H} =  e^{-\frac{1}{\eta} \left( \beta\left( e^{x} + e^y\right)  - \omega  y \right)} . \nonumber %\label{Eq:HamiltonY}
\end{eqnarray} 

The maximum of this probability corresponds to the minima of the Hamiltonian (\ref{Eq:DefH}). This Hamiltonian, $H$, has no minima in the $x$-direction, thus the most probable situation is as $x\to -\infty$, that is $\lambda \to 0$. On the other hand, $H$ has a  minimum along the $y$ direction, which is for $e^y=\nu$, that is $\sigma = \nu$. Therefore, we conclude that in the stochastic model the social system presents a small probability of damaging events, while the latent population to engage in disorder and join riots is in average critical $\sigma = \sigma_c=\nu$. Because of this fact, the model presents similarities with the mechanism of self-organized criticality \cite{soc}, which is common in many ``cathastrophic'' systems, such as avalanches and earthquakes \cite{bak}, and is also present in epidemic diseases\cite{Rhodes1997}.

\section{Data Analysis}
\label{Sec:DataAnalysis}

To fit the data with the solution of the system of o.d.e.'s (\ref{Eq:lambda}) and (\ref{Eq:sigma}), one needs to fit the parameters $\omega,\beta $ together with an initial condition, $\{t_0, \lambda_0 ,\sigma_0\}$; that is, we need to find the best fit varying five parameters. As a first guess, we can estimate some of these parameters from the data and the analytic results of Section \ref{Sec:Epidemic}. First, the asymptotic expression (\ref{Eq:AsymptoticSolution}) allows us to estimate $\omega$ from a simple linear regression that fits the observations in Fig. \ref{Fig:DataHRSlopes}.

\begin{figure}[h!]
\begin{center}
 \includegraphics[width=7cm]{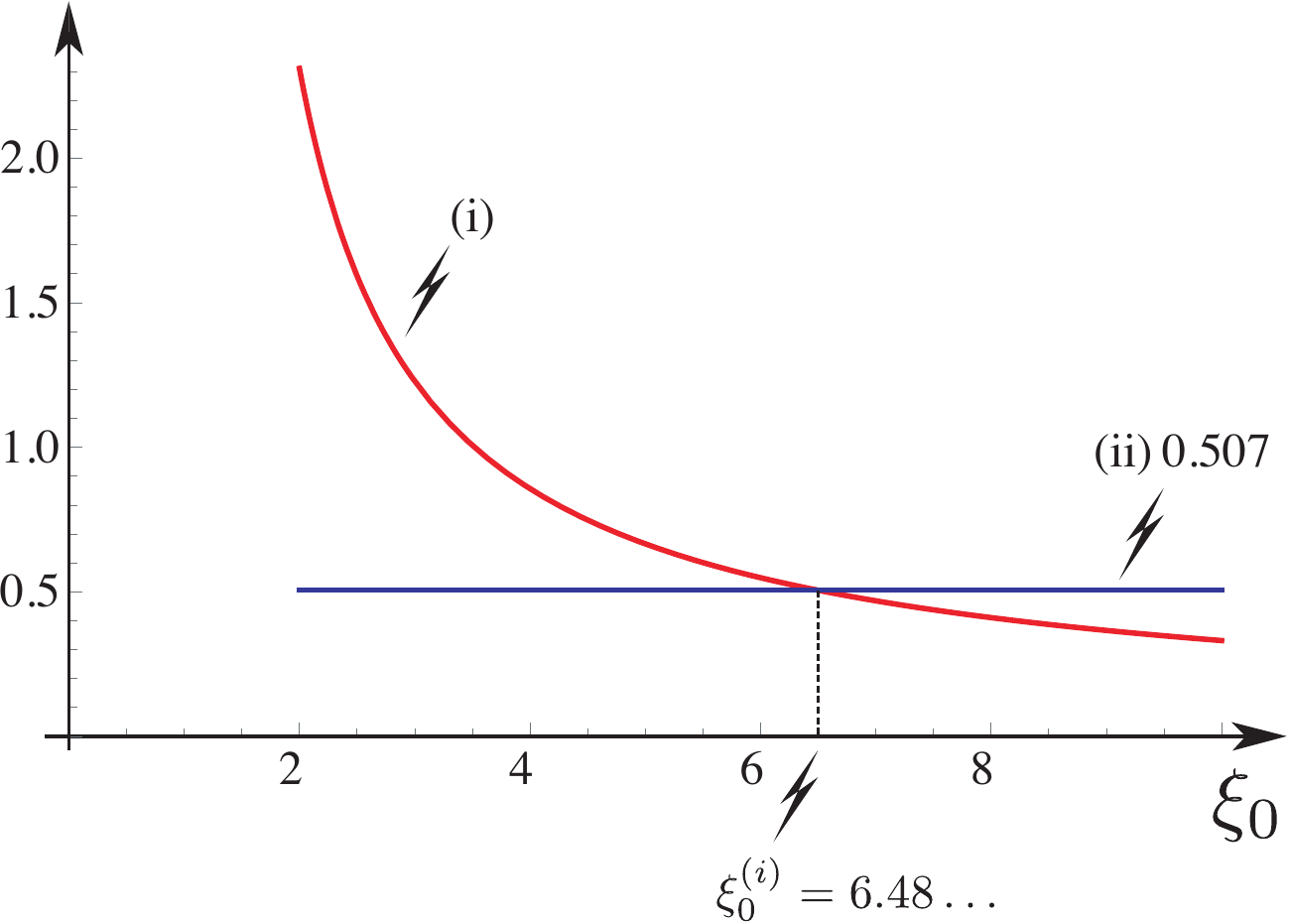}
 \end{center}
\caption{ \label{Fig:EqIntegral5}  Plot of both, (i) the l.h.s and (ii) the r.h.s of (\ref{Eq:Integral5}) as a function of $\xi_0$ for $\lambda_0=100$[e], $\lambda_{max}=350 $[e], $t_0=2$[d], $t_{max}=3$[d], and $\omega^{(i)}= 0.5065 $[1/d]. The value of  $ \omega (t_{\rm max}-t_0)= 0.5065$, and the intersection provides the solution for $\xi_0^{(i)} = 6.4847$. Finally, from  (\ref{Eq:Integral4} ) one computes $\nu = 69.15 $[e].}
\end{figure}

Next, imposing directly from the data that at $t=t_0$ the number of events is $\lambda_0$, then the remaining two parameters $\beta$ and $\sigma_0$ follow after imposing  that at $t=t_{\rm max}$ the number of events reaches its maximum value of $\lambda_{\rm max}=\lambda(t_{\rm max})$, for which $\sigma =\nu$ (see previous Section \ref{SubSecc:Activation}). Thus after (\ref{Eq:Integral2}) one gets: 
\begin{eqnarray}
\int_{1}^{\xi_0} \frac{d\xi}{\xi \left( { \lambda_0}/{\nu}+ \log\left({ \xi}/{\xi_0} \right)  - (\xi -\xi_0)   \right)} &=& { \omega}(t_{\rm max}-t_0).
\label{Eq:Integral3} 
\end{eqnarray}
Here we  have used the change of variables $\xi = \sigma/\nu$ and we denote $\xi_0 = {\sigma_0}/{\nu} \equiv \mathcal{R}_0$ for the reproduction number.
%\begin{eqnarray}\int_{\sigma_0}^{\nu} \frac{d\sigma}{\sigma \left(  \lambda_0+ \nu\log\left(\frac{ \sigma}{\sigma_0} \right)  - (\sigma -\sigma_0)   \right)} &=&  -\frac{1}{\nu} \omega (t_{\rm max}-t_0),\label{Eq:Integral3} 
%\end{eqnarray}
Eq. (\ref{Eq:Integral3})  provides a first relation among  $\xi_0 $ and $\nu$ for a given set of parameters $(t_0, t_{\rm max}, \lambda_0, \lambda_{\rm max})$.
The second relation comes from (\ref{Eq:Integral0}), namely:
\begin{eqnarray}
\frac{1}{\nu} \left(  \lambda_{max} - \lambda_0\right)&=&    \xi_0- \log{\xi_0}   -1 .  \label{Eq:Integral4} 
\end{eqnarray}
%Therefore, among the five parameters only $\beta$ and $\sigma_0$ need to be calibrated to find the best fit. 
Computing $\nu$ from (\ref{Eq:Integral4}) and introducing it into equation (\ref{Eq:Integral3}), one gets  relation for $\xi_0$:
 \begin{widetext}
\begin{eqnarray}
\int_{1}^{\xi_0} \frac{d\xi}{\xi \left(\frac{ \lambda_0}{ \lambda_{max} - \lambda_0} \left(  \xi_0- \log{\xi_0}   -1\right) + \log\left({ \xi}/{\xi_0} \right)  - (\xi -\xi_0)   \right)} &=& { \omega}(t_{\rm max}-t_0),
\label{Eq:Integral5} 
\end{eqnarray}
 \end{widetext}
 as an implicit function of all other parameters, $(t_0, t_{\rm max}, \lambda_0, \lambda_{\rm max})$. To solve the equation (\ref{Eq:Integral5}) for $\xi_0$, we may use a straight-forward graphic procedure: for given values of $\lambda_{0}$ and $\lambda_{max}$ one plots the left-hand side of (\ref{Eq:Integral5}) as a function of $\xi_0$, then the intersection of this curve with the constant given value ${ \omega}(t_{\rm max}-t_0)$ provides directly $\xi_0$. Finally, one completes the original system (\ref{Eq:Integral3},\ref{Eq:Integral4}), computing $\nu$ from  (\ref{Eq:Integral4}). Fig. \ref{Fig:EqIntegral5} illustrates the procedure for the first peak of Fig. \ref{Fig:DataHR}.

Repeating the same procedure for each event, one can obtain suitable parameters to characterize them. 
The parameters, $(t^{(i)}_0, \lambda^{(i)}_0,\xi^{(i)}_0, \omega^{(i)},\nu^{(i)} )$, obtained by this method are not necessarily the best set of parameters to fit the data by the solution of the set of ordinary differential equations (\ref{Eq:lambda}) and (\ref{Eq:sigma}) over the time interval of the event duration. To quantify the precision of the realized model we compute the mean squared error:
$$ {\rm error} =\frac{1}{K} \sum_{n=0}^{K-1} \left( \frac{\lambda(t_n)}{\lambda_{n}} -1 \right)^2, $$ where $\lambda(t)$ is the numerical solution of the o.d.e. system (\ref{Eq:lambda}) and (\ref{Eq:sigma}) using  the obtained parameters, thus, $\lambda(t_n)$ is this solution evaluated at $t_n$. On the other hand, $\lambda_{n}$ is the value from the data at day $t_n$.
 
To improve the initial estimations, we use the previous calculation, $(t^{(i)}_0, \lambda^{(i)}_0,\xi^{(i)}_0, \omega^{(i)},\nu^{(i)} )$, as a starting guess, then  we explore randomly $10^4$ trials around this seed point for a smaller error; that is, better set of parameters. More precisely,  the initial guess point belongs to a space in five dimensions and the exploration is bounded to a vicinity of linear size $\Delta$. A typical neighborhood is defined for instance by: $\lambda_0  \in [ \lambda^{(i)}_0 (1-\Delta/2), \lambda^{(i)}_0 (1+\Delta/2) ]$ and the same for other parameters. Finally, this random search is iterated and the vicinity size is reduced by half ($\Delta\to \Delta/2$) at each step. For all cases after 5 iterations, the procedure converges to a better set of parameters, improving the original estimation by a factor of between 10 and 100  (see Table \ref{Table}).  The final set of parameters will be denoted by: $(t^{*}_0, \lambda^{*}_0,\xi^{*}_0, \omega^{*},\nu^{*} )$.  

Table \ref{Table} summarizes the initial guesses from the conditions (\ref{Eq:Integral4}) and  (\ref{Eq:Integral5}) and the final values after convergence. Lastly, the mean squared error criteria of convergence is also shown. 

\begin{figure}[h!]
\begin{center}
 \includegraphics[width=7cm]{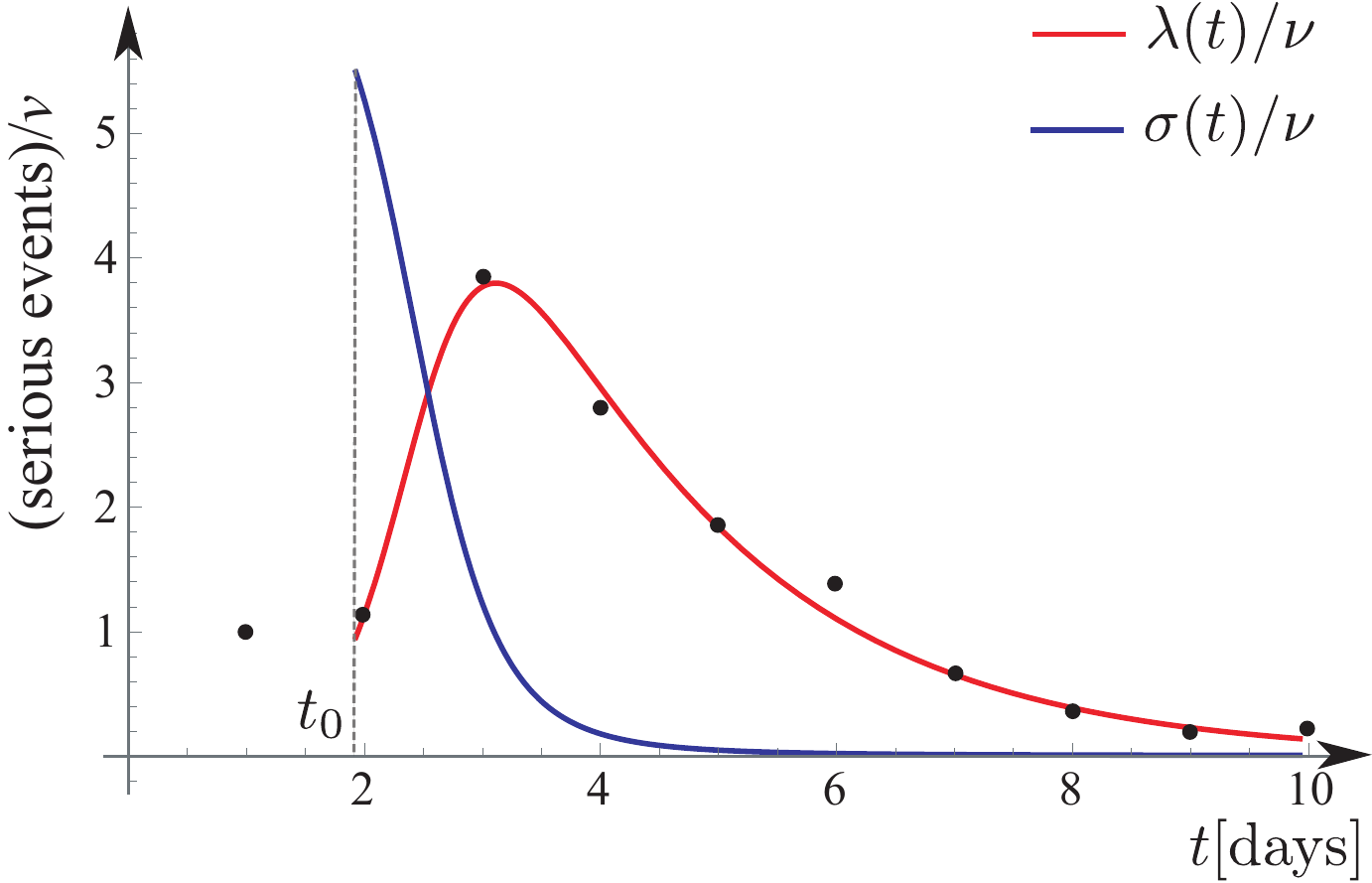} 
 \end{center}
\caption{\label{Fig:Event1} Plot of the normalized riot activity $\lambda(t)/\nu$ and the potential rioters $\sigma(t)/\nu$ as a function of time. The data is also normalized by $\nu$. For the simulation, the parameters are $ t_0 = 1.927$, $\lambda_0 =87.37$,  $\sigma_0 = 489.2$, $\omega = 0.521$, $\beta =0.00626$, and $\nu= 83.169$.  }
\end{figure}

As an example, we provide in Fig. \ref{Fig:Event1} the best fitting found for the case of the first and largest event shown in Fig. \ref{Fig:DataHR}.

 %\begin{widetext}
\begin{table*}[h!]
 \caption{\label{Table} Fitting parameters for eight events from October 18$^{th}$, 2019 up to March 5$^{th}$, 2020. }
\begin{tabular}{||c||c|c|c|c|c|c|c|c||c|c|c|c|c|c|c|c||}\hline
peak & $t^{(i)}_0$ \, [d] & $t^{(i)}_{\rm max}$ & $\lambda^{(i)}_0$ [e]  & $\lambda^{(i)}_{\rm max}$ & $\xi^{(i)}_0$ &  $\omega^{(i)}$ \, [1/d ]  & $\nu^{(i)}  $ [e] & error & $t^*_0$ \, [d] & $t^*_{\rm max}$ & $\lambda^*_0$ [e]  & $\lambda^*_{\rm max}$ & $\xi^*_0$ &  $\omega^*$ \, [1/d ]  & $\nu^* $ [e] & error \\ % & date for $t_0$ \\
\hline
1 &2& 3  &100 & 350&  6.484& 0.507& 69.15& 0.0353& 1.927& 3.145 &87.37& 345.24 & 5.882&0.521& 83.169& 0.0263 %& Oct 19$^{\rm th}$, 2019\\
\\ \hline
2 &10 & 11& 21 & 89& 4.666& 0.804 & 31.99 & 0.222& 9.926& 10.866 &17.922& 71.404& 4.501&0.847&27.591&0.0515 %Oct 28$^{\rm th}$, 2019 
\\
\hline
3 & 19& 20& 26& 84& 5.486& 0.572& 20.83& 0.499 & 18.822 & 19.811 & 21.606 & 67.59  &5.256 & 0.596 &17.732&0.054 \\
\hline
4 &25& 26& 50& 189& 5.527& 0.627& 49.33& 2.107  & 24.866 &25.724 & 43.811& 135.77 &4.775&0.717&41.558&0.134 % Nov 12$^{\rm th}$, 2019 
\\
\hline
5 & 34& 35& 16& 86& 4.932& 0.849& 29.96& 0.534 & 33.961& 34.87& 15.105& 74.83& 4.834&0.997&26.371& 0.0298  %Nov 21$^{\rm st}$, 2019  
\\
\hline
6 &38& 40& 7& 99& 4.345& 0.684& 49.05& 0.988 & 37.958 & 39.575 & 6.189 & 85.03 & 4.377&0.814&40.79&0.039  %Nov 25$^{\rm th}$, 2019  
 \\
\hline
7&103& 104& 2& 30& 8.03& 0.715& 5.660& 0.0447 & 102.99&104.003 & 1.741 & 30.06 &7.905 &0.768 &5.884 &0.0212 %Jan 29$^{\rm th}$, 2020  
\\
\hline
8 &136& 137& 7& 28& 5.027& 0.717& 8.705 & 0.700 &  135.94 & 136.8 &6.493&21.97 & 4.640 & 0.843 & 7.334 &0.045  %Mar 4$^{\rm th}$, 2020 
\\
\hline
\end{tabular}
\end{table*}
 %\end{widetext}

From the data summarized in Table \ref{Table}, some interesting features can be observed: Among all figures, the values of $\xi_0\equiv \sigma_0 /\nu$ are more or less uniform for all cases. On average one notices that  $\left< \sigma_0 /\nu\right> \approx 5.25$, indicating that all events have almost the same reproduction number with value greater than one, and hence the epidemics take off with almost the same strength. Similarly, although the value of the maximum number of events is quite disparate, the ratio  $\lambda_{\rm max} /\nu$ also varies among a well-defined value around   $\left<\lambda_{\rm max} /\nu\right>  \approx 3.34$. This is not surprising, in light of the relation (\ref{Eq:Integral4}) that provides $\lambda_{\rm max}/\nu$ as a single function of $\sigma_0 /\nu$ if $\lambda_0\ll \lambda_{\rm max}$. 
  Because $\xi_0= \sigma_0 /\nu$ is not close to unity, it is expected that the event peaks manifest characteristic asymmetry respect to the peak maxima, as already noticed in Section \ref{SubSec:Empirical}. Indeed, the taking-off phase of an event is sharper than the exponential decaying stage after the event. Symmetric peaks are a result valid if $\sigma_0 \gtrsim \nu$, in which case it is legitimate to approximate the denominator in equation (\ref{Eq:Integral2}) by a quadratic form.     
  
Another remarkable feature of the anatomy of the Chilean episodes is that the time scale $\omega$ does not vary substantially across all observed events. On average, $\left<\omega\right> \approx 0.76 \, [1/{\rm d}]$; that is, the mean lifetime is of the order of one day and eight hours. More precisely, after two days, the activity decreases by a factor of 5 from its peak, and after three days, the event magnitude decreases by a factor of 10. 
Although the US disturbances\cite{Burbeck1978} and the French riots\cite{bonnasse2018epidemiological} share many similarities with the Chilean unrest phenomena - in particular with respect  to the temporal scales ($1\lesssim \omega \lesssim 3 $ [1/d] for the USA, $\omega\approx 0.25$ [1/d] for France, and $\omega \approx 0.75$ [1/d] for Chile) -  we emphasize that the Chilean case is the only one that shows sizeable rebounds and sequels, which, to our knowledge, is unique and makes this historic episode all the more remarkable. The sequels present a regular periodicity of event occurrence times across the first six events, with the interval sequence between consecutive $t_{\rm max}$ being $\{8, 9, 6, 9, 4\} $ days. This suggests that the Chilean riots might be interpreted in terms of periodic external forcing, as discussed in Section \ref{SubSec:forcing}.

\section{Discussion}
\label{Sec:Discussion}

Motivated by the 2019 Chilean social unrest that caused costly infrastructure damage and serious injuries to a group of the population, we take advantage of mathematical modelling and numerical simulations to analyze this complex social phenomenon and compare findings with public aggregated data, thereby increasing our understanding of how this type of social event evolves through time.

 We analyze the dynamics of the Chilean rioting episodes that started in October 2019 from the perspective of epidemic-like models.  
We show that this approach can be interpreted in a relatively easy way and could be used to understand the factors that determine the dynamics of rioting events, and qualitatively guide policymakers to make informed decisions aimed at ensuring public safety. In addition, the approach offers a richer explanatory power due to the existence of an underlying Hamiltonian structure. In fact, under the influence of a periodic forcing, the dynamics may display three types of behavior depending on the parameters: 1) periodic oscillation; 2) quasi-periodic oscillation; or 3) chaotic dynamics. On the other hand, in the presence of a stochastic forcing, the most probable behavior places the system in a critical condition for contagion. Interestingly, this result tells us that social turmoil events may be in the same class as the well-known phenomena of self-organized criticality; something that we plan to study in the future.   

Although all these properties are exciting and promising, it seems clear that the observed Chilean social phenomenon requires a more careful and interdisciplinary effort. In particular, several questions remain open: From a psychological perspective, what triggered the rage to flourish in a substantial percentage of the population, provoking massive sentiments of unfairness that found sympathy even in the well-off social class? How could we include these psychological factors in an epidemic-like model? How did the riot events spread spatially, stimulating episodes of disorder throughout the country? From a social network perspective, can we understand the speed at which ideas and opinions (fake and true) become popular in social media, influencing group thinking and, therefore, group behavior? What enables the spread of fake news, and how damaging are they in terms of group contagion that can lead to catastrophic outcomes? How can we avoid the spread of these fake news within a social network? These last three questions can readily be framed within our modelling approach, since they relate to the structure and rate of contagion.

It should be emphasized that modelling efforts such as those presented here are strictly complementary to other avenues via which social disorder can be prevented. As outlined in the paper, riots have complex sociological causes - often grounded in hardship - and any strategy or policy must necessarily include aspects which address these and reduce the motivation to engage in disorder. Nevertheless, such factors are typically chronic and can only be addressed in the long term; furthermore, it would be na\"ive to expect that the motivation to riot could ever be eradicated. Given this, the planning and testing of practical responses is a necessary component of any strategy, in order to minimize the negative consequences of disorder when it does occur. If mathematical modelling and numerical simulations are able to describe and predict the dynamics of violent social turmoils, they become a powerful tool to provide authorities involved in politics and security with useful insights to understand this complex social phenomena. This information can be used as an input to make informed decisions and attempt to prevent contagion effects of rioting.



\vskip 1 cm

\noindent{\bf AUTHOR'S CONTRIBUTIONS}
\medskip

J.O. conducted the socio-economical study, and  wrote the Introduction, Section \ref{Sec:Chile} and the Appendix \ref{Sec:timeline}. S.R. and K.V. developed the mathematical modeling (Section \ref{Sec:Epidemic}). P.C. and S.R. realized the data analysis \ref{Sec:DataAnalysis}. All the authors have contributed in the discussion Section \ref{Sec:Discussion} and read and approved the manuscript.

\begin{acknowledgments}
We acknowledge that initial thoughts on this work occurred during discussions with P. Montebruno.  P.C., J.O., S.R., and K.V. thank the  Facultad de Ingenier\'ia y Ciencias (Universidad Adolfo Ib\'a\~nez)  for the research fund provided to carry out this research, and C.C. wishes to acknowledge the support of Universidad de los Andes (CL) through FAI initiatives.
\end{acknowledgments}

\vskip 1 cm 

\noindent{\bf Data Availability Statement}
\medskip
The data that supports the findings of this study are openly available in the Undersecretary for Human Rights' web page at http://ddhh.minjusticia.gob.cl/informacion-sobre-la-situacion-del-pais-desde-el-19-de-octubre, reference number: ~\cite{hhrr}.


\section{Appendixes}

\appendix
\section{Timeline summary}
\label{Sec:timeline}

This Appendix summarizes the main events of the social unrest discussed in Section \ref{Sec:Context} and Fig. \ref{Fig:TimeLine}.

\begin{itemize}

\item Monday { October} 7$^{\rm th}$, 2019: Under the slogan \textit{``Evade!''} secondary students self-organized via social networks to massively evade the subway fare in the capital city of Santiago as a consequence of a 3.75 percent fare hike of CLP\$30. Although this increase represents less than 5 cents of a US dollar, low-income families spend between 13 and 28 percent of their budget on public transportation.

\item Monday { October} 14$^{\rm th}$, 2019: Student protests continue in Santiago and several stations on Line 5 were closed in the afternoon after violent incidents were reported. 

\item Friday  { October} 18$^{\rm th}$, 2019 (The zero day): The escalation of protests. Within two weeks the social unrest turned massive and violent; destruction of the subway stations in Santiago was coordinated, damaging almost 60 percent of them. The entire Metro system was closed after the attacks, and it remained closed for several days and weeks. The police force tried dispersing the crowds firing rubber bullets, teargas and water cannons.  

\item Saturday { October} 19$^{\rm th}$, 2019: The violence continues. Shops were looted, buses were set alight and clashes occurred between the security forces and rioters. The President imposed a curfew in Santiago from 22:00 to 07:00 hours and the army was taken to the streets. After rioters started spreading to other cities, the President declared a state of emergency in Valapara\'iso and Concepci\'on. The increase in the metro fare was cancelled. 

\item Sunday { October} 20$^{\rm th}$, 2019: As violent protests continued, the President publicly announced on the evening that \textit{``We are at war against a powerful enemy...''}. \footnote{``Estamos en guerra contra un enemigo poderoso, implacable, que no respeta a nada ni a nadie y que est\'a dispuesto a usar la violencia y la delincuencia sin ning\'un l\'imite''. President Sebasti\'an Pi\~nera in press conference on Sunday Oct. 20$^{\rm th}$, 2019.} The next day, the head of national defense, Javier Iturriaga, assured in a press conference that \textit{``I am not at war with anyone..."}. As violent protests continued spreading, curfews were imposed in the Santiago Metropolitan Region, and the regions of Valpara\'iso, B\'iob\'io, and Coquimbo.   %en respuesta a los diferentes actos de violencia que se registraron en el país.

\item Friday { October} 25$^{\rm th}$, 2019: In spite of the palliative measures promoted by the Chilean Government, the citizens went back to the streets. Over a million people took to the streets in Santiago, and thousands more throughout Chile, becoming the largest pacific demonstration after Chile returned to democracy. 

\item Wednesday { October} 30$^{\rm th}$, 2019: Amid the ongoing clashes between protesters and security forces, the Chilean Government decides to pull out of hosting the Asia-Pacific Economic Cooperation (APEC) and The UN Climate Change Conference (COP 25). 

\item Sunday { November} 10$^{\rm th}$, 2019: Towards a new Constitution. The Government agreed to initiate a process to draft a new Constitution.

\item Monday { November} 12$^{\rm th}$, 2019: Change of cabinet. As a result of the massive protests, the President reshuffled his cabinet. He replaced the interior and finance ministers, and the government spokesperson.

\item Friday { November} 15$^{\rm th}$, 2019: Lawmakers agreed and signed an ``Agreement for Peace and a New Constitution'', an agreement that was regarded as ``historic''. In a referendum, initially scheduled for April 26$^{\rm th}$ 2020 but postponed to October 2020 due to the Covid-19 pandemia, voters will be asked whether they approve the idea of a new constitution and whether current lawmakers should serve on the commission that would redraft the document.

\item Tuesday { November} 19$^{\rm th}$, 2019: The Inter-American Court of Human Rights, an agency under the Organization of American States, condemned the excessive use of force during social protests and called on the authorities to order state security forces to immediately cease their disproportionate use of force. Amid public outcry over eye injuries suffered by hundreds of protesters, the police decides to cease the use of rubber bullets against rioters. %, given the questions that these types of bullets have received.

\item Friday { November} 22$^{nd}$, 2019: Rejection of the Amnesty International report. Both the Government and the Chilean Army openly rejected the report prepared by Amnesty International, a non-governmental organization that investigates violations of human rights occurred in different countries. The organization recorded at least 23 cases of human rights violations, documented in eight different regions since the start of the social unrest.

\item Saturday { November} 23$^{rd}$, 2019: The social turmoil leaves 23 dead. After the President recognized that there may have been a breach of protocols for the use of force by the police in Chile, the official death toll was announced during the social outbreak, which reached 23.

\end{itemize}


\begin{thebibliography}{41}%
\makeatletter
\providecommand \@ifxundefined [1]{%
 \@ifx{#1\undefined}
}%
\providecommand \@ifnum [1]{%
 \ifnum #1\expandafter \@firstoftwo
 \else \expandafter \@secondoftwo
 \fi
}%
\providecommand \@ifx [1]{%
 \ifx #1\expandafter \@firstoftwo
 \else \expandafter \@secondoftwo
 \fi
}%
\providecommand \natexlab [1]{#1}%
\providecommand \enquote  [1]{``#1''}%
\providecommand \bibnamefont  [1]{#1}%
\providecommand \bibfnamefont [1]{#1}%
\providecommand \citenamefont [1]{#1}%
\providecommand \href@noop [0]{\@secondoftwo}%
\providecommand \href [0]{\begingroup \@sanitize@url \@href}%
\providecommand \@href[1]{\@@startlink{#1}\@@href}%
\providecommand \@@href[1]{\endgroup#1\@@endlink}%
\providecommand \@sanitize@url [0]{\catcode `\\12\catcode `\$12\catcode
  `\&12\catcode `\#12\catcode `\^12\catcode `\_12\catcode `\%12\relax}%
\providecommand \@@startlink[1]{}%
\providecommand \@@endlink[0]{}%
\providecommand \url  [0]{\begingroup\@sanitize@url \@url }%
\providecommand \@url [1]{\endgroup\@href {#1}{\urlprefix }}%
\providecommand \urlprefix  [0]{URL }%
\providecommand \Eprint [0]{\href }%
\providecommand \doibase [0]{http://dx.doi.org/}%
\providecommand \selectlanguage [0]{\@gobble}%
\providecommand \bibinfo  [0]{\@secondoftwo}%
\providecommand \bibfield  [0]{\@secondoftwo}%
\providecommand \translation [1]{[#1]}%
\providecommand \BibitemOpen [0]{}%
\providecommand \bibitemStop [0]{}%
\providecommand \bibitemNoStop [0]{.\EOS\space}%
\providecommand \EOS [0]{\spacefactor3000\relax}%
\providecommand \BibitemShut  [1]{\csname bibitem#1\endcsname}%
\let\auto@bib@innerbib\@empty
%</preamble>
\bibitem [{\citenamefont {{Verisk Maplecroft}}(2020)}]{Verisk}%
  \BibitemOpen
  \bibfield  {author} {\bibinfo {author} {\bibnamefont {{Verisk Maplecroft}}},\
  }\href@noop {} {\enquote {\bibinfo {title} {{Political Risk Outlook 2020}},}\
  }\bibinfo {howpublished}
  {{https://www.maplecroft.com/insights/analysis/47-countries-witness-surge-in-civil-unrest/}}
  (\bibinfo {year} {2020})\BibitemShut {NoStop}%
\bibitem [{\citenamefont {Spilerman}(1970)}]{spilerman1970causes}%
  \BibitemOpen
  \bibfield  {author} {\bibinfo {author} {\bibfnamefont {S.}~\bibnamefont
  {Spilerman}},\ }\bibfield  {title} {\enquote {\bibinfo {title} {The causes of
  racial disturbances: {A} comparison of alternative explanations},}\
  }\href@noop {} {\bibfield  {journal} {\bibinfo  {journal} {American
  Sociological Review}\ ,\ \bibinfo {pages} {627--649}} (\bibinfo {year}
  {1970})}\BibitemShut {NoStop}%
\bibitem [{\citenamefont {Burbeck}, \citenamefont {Raine},\ and\ \citenamefont
  {Abudu~Stark}(1978)}]{Burbeck1978}%
  \BibitemOpen
  \bibfield  {author} {\bibinfo {author} {\bibfnamefont {S.~L.}\ \bibnamefont
  {Burbeck}}, \bibinfo {author} {\bibfnamefont {W.~J.}\ \bibnamefont {Raine}},
  \ and\ \bibinfo {author} {\bibfnamefont {M.~J.}\ \bibnamefont
  {Abudu~Stark}},\ }\bibfield  {title} {\enquote {\bibinfo {title} {The
  dynamics of riot growth: An epidemiological approach},}\ }\href {\doibase
  10.1080/0022250X.1978.9989878} {\bibfield  {journal} {\bibinfo  {journal}
  {The Journal of Mathematical Sociology}\ }\textbf {\bibinfo {volume} {6}},\
  \bibinfo {pages} {1--22} (\bibinfo {year} {1978})}\BibitemShut {NoStop}%
\bibitem [{\citenamefont {Granovetter}(1978)}]{Granovetter}%
  \BibitemOpen
  \bibfield  {author} {\bibinfo {author} {\bibfnamefont {M.}~\bibnamefont
  {Granovetter}},\ }\bibfield  {title} {\enquote {\bibinfo {title} {Threshold
  models of collective behavior},}\ }\href
  {http://www.jstor.org/stable/2778111} {\bibfield  {journal} {\bibinfo
  {journal} {American Journal of Sociology}\ }\textbf {\bibinfo {volume}
  {83}},\ \bibinfo {pages} {1420--1443} (\bibinfo {year} {1978})}\BibitemShut
  {NoStop}%
\bibitem [{\citenamefont {Berestycki}, \citenamefont {Nadal},\ and\
  \citenamefont {Rodr\'iguez}(2015)}]{berestycki2015model}%
  \BibitemOpen
  \bibfield  {author} {\bibinfo {author} {\bibfnamefont {H.}~\bibnamefont
  {Berestycki}}, \bibinfo {author} {\bibfnamefont {J.-P.}\ \bibnamefont
  {Nadal}}, \ and\ \bibinfo {author} {\bibfnamefont {N.}~\bibnamefont
  {Rodr\'iguez}},\ }\bibfield  {title} {\enquote {\bibinfo {title} {A model of
  riots dynamics: Shocks, diffusion and thresholds},}\ }\href {\doibase
  10.3934/nhm.2015.10.443} {\bibfield  {journal} {\bibinfo  {journal} {Networks
  \& Heterogeneous Media}\ }\textbf {\bibinfo {volume} {10}},\ \bibinfo {pages}
  {443--475} (\bibinfo {year} {2015})}\BibitemShut {NoStop}%
\bibitem [{\citenamefont {Baudains}, \citenamefont {Braithwaite},\ and\
  \citenamefont {Johnson}(2016)}]{Baudains2016london}%
  \BibitemOpen
  \bibfield  {author} {\bibinfo {author} {\bibfnamefont {P.}~\bibnamefont
  {Baudains}}, \bibinfo {author} {\bibfnamefont {A.}~\bibnamefont
  {Braithwaite}}, \ and\ \bibinfo {author} {\bibfnamefont {S.~D.}\ \bibnamefont
  {Johnson}},\ }\bibfield  {title} {\enquote {\bibinfo {title} {The {L}ondon
  {R}iots -- 2: {A} {D}iscrete {C}hoice {M}odel},}\ }in\ \href@noop {} {\emph
  {\bibinfo {booktitle} {Approaches to Geo‐mathematical Modelling: New Tools
  for Complexity Science}}}\ (\bibinfo  {publisher} {Wiley Online Library},\
  \bibinfo {year} {2016})\ pp.\ \bibinfo {pages} {170--191}\BibitemShut
  {NoStop}%
\bibitem [{\citenamefont {Bonnasse-Gahot}\ \emph {et~al.}(2018)\citenamefont
  {Bonnasse-Gahot}, \citenamefont {Berestycki}, \citenamefont {Depuiset},
  \citenamefont {Gordon}, \citenamefont {Roch{\'e}}, \citenamefont
  {Rodriguez},\ and\ \citenamefont {Nadal}}]{bonnasse2018epidemiological}%
  \BibitemOpen
  \bibfield  {author} {\bibinfo {author} {\bibfnamefont {L.}~\bibnamefont
  {Bonnasse-Gahot}}, \bibinfo {author} {\bibfnamefont {H.}~\bibnamefont
  {Berestycki}}, \bibinfo {author} {\bibfnamefont {M.-A.}\ \bibnamefont
  {Depuiset}}, \bibinfo {author} {\bibfnamefont {M.~B.}\ \bibnamefont
  {Gordon}}, \bibinfo {author} {\bibfnamefont {S.}~\bibnamefont {Roch{\'e}}},
  \bibinfo {author} {\bibfnamefont {N.}~\bibnamefont {Rodriguez}}, \ and\
  \bibinfo {author} {\bibfnamefont {J.-P.}\ \bibnamefont {Nadal}},\ }\bibfield
  {title} {\enquote {\bibinfo {title} {Epidemiological modelling of the 2005
  {F}rench riots: a spreading wave and the role of contagion},}\ }\href@noop {}
  {\bibfield  {journal} {\bibinfo  {journal} {Scientific reports}\ }\textbf
  {\bibinfo {volume} {8}},\ \bibinfo {pages} {107} (\bibinfo {year}
  {2018})}\BibitemShut {NoStop}%
\bibitem [{\citenamefont {Salehyan}\ \emph {et~al.}(2012)\citenamefont
  {Salehyan}, \citenamefont {Hendrix}, \citenamefont {Hamner}, \citenamefont
  {Case}, \citenamefont {Linebarger}, \citenamefont {Stull},\ and\
  \citenamefont {Williams}}]{salehyan2012social}%
  \BibitemOpen
  \bibfield  {author} {\bibinfo {author} {\bibfnamefont {I.}~\bibnamefont
  {Salehyan}}, \bibinfo {author} {\bibfnamefont {C.~S.}\ \bibnamefont
  {Hendrix}}, \bibinfo {author} {\bibfnamefont {J.}~\bibnamefont {Hamner}},
  \bibinfo {author} {\bibfnamefont {C.}~\bibnamefont {Case}}, \bibinfo {author}
  {\bibfnamefont {C.}~\bibnamefont {Linebarger}}, \bibinfo {author}
  {\bibfnamefont {E.}~\bibnamefont {Stull}}, \ and\ \bibinfo {author}
  {\bibfnamefont {J.}~\bibnamefont {Williams}},\ }\bibfield  {title} {\enquote
  {\bibinfo {title} {Social conflict in {A}frica: {A} new database},}\
  }\href@noop {} {\bibfield  {journal} {\bibinfo  {journal} {International
  Interactions}\ }\textbf {\bibinfo {volume} {38}},\ \bibinfo {pages}
  {503--511} (\bibinfo {year} {2012})}\BibitemShut {NoStop}%
\bibitem [{\citenamefont {Davies}\ \emph {et~al.}(2013)\citenamefont {Davies},
  \citenamefont {Fry}, \citenamefont {Wilson},\ and\ \citenamefont
  {Bishop}}]{Davies2013}%
  \BibitemOpen
  \bibfield  {author} {\bibinfo {author} {\bibfnamefont {T.~P.}\ \bibnamefont
  {Davies}}, \bibinfo {author} {\bibfnamefont {H.~M.}\ \bibnamefont {Fry}},
  \bibinfo {author} {\bibfnamefont {A.~G.}\ \bibnamefont {Wilson}}, \ and\
  \bibinfo {author} {\bibfnamefont {S.~R.}\ \bibnamefont {Bishop}},\ }\bibfield
   {title} {\enquote {\bibinfo {title} {{A mathematical model of the {L}ondon
  riots and their policing}},}\ }\href {\doibase 10.1038/srep01303} {\bibfield
  {journal} {\bibinfo  {journal} {Scientific Reports}\ }\textbf {\bibinfo
  {volume} {3}} (\bibinfo {year} {2013}),\ 10.1038/srep01303}\BibitemShut
  {NoStop}%
\bibitem [{\citenamefont {Davies}\ and\ \citenamefont
  {Marchione}(2015)}]{davies2015event}%
  \BibitemOpen
  \bibfield  {author} {\bibinfo {author} {\bibfnamefont {T.}~\bibnamefont
  {Davies}}\ and\ \bibinfo {author} {\bibfnamefont {E.}~\bibnamefont
  {Marchione}},\ }\bibfield  {title} {\enquote {\bibinfo {title} {Event
  networks and the identification of crime pattern motifs},}\ }\href@noop {}
  {\bibfield  {journal} {\bibinfo  {journal} {PloS one}\ }\textbf {\bibinfo
  {volume} {10}},\ \bibinfo {pages} {e0143638} (\bibinfo {year}
  {2015})}\BibitemShut {NoStop}%
\bibitem [{\citenamefont {Ffrench-{D}avis}(2016)}]{fd2014}%
  \BibitemOpen
  \bibfield  {author} {\bibinfo {author} {\bibfnamefont {R.}~\bibnamefont
  {Ffrench-{D}avis}},\ }\href {\doibase
  {https://doi.org/https://doi.org/10.18356/fd11329b-en}} {\emph {\bibinfo
  {title} {Is {C}hile a role model for development?, in Calcagno, A., et al.
  (eds.), {R}ethinking {D}evelopment {S}trategies after the {F}inancial
  {C}risis: Volume II - {C}ountry {S}tudies and {I}nternational
  {C}omparisons}}}\ (\bibinfo  {publisher} {UN, New York},\ \bibinfo {year}
  {2016})\BibitemShut {NoStop}%
\bibitem [{\citenamefont {{The {W}orld {B}ank, {D}evelopment {R}esearch
  {G}roup}}()}]{wbdata}%
  \BibitemOpen
  \bibfield  {author} {\bibinfo {author} {\bibnamefont {{The {W}orld {B}ank,
  {D}evelopment {R}esearch {G}roup}}},\ }\href@noop {} {\enquote {\bibinfo
  {title} {{G}lobal {D}evelopment data},}\ }\bibinfo {note}
  {{https://data.worldbank.org/}}\BibitemShut {NoStop}%
\bibitem [{\citenamefont {Flores}\ \emph {et~al.}(2019)\citenamefont {Flores},
  \citenamefont {Sanhueza}, \citenamefont {Atria},\ and\ \citenamefont
  {Mayer}}]{flores2019top}%
  \BibitemOpen
  \bibfield  {author} {\bibinfo {author} {\bibfnamefont {I.}~\bibnamefont
  {Flores}}, \bibinfo {author} {\bibfnamefont {C.}~\bibnamefont {Sanhueza}},
  \bibinfo {author} {\bibfnamefont {J.}~\bibnamefont {Atria}}, \ and\ \bibinfo
  {author} {\bibfnamefont {R.}~\bibnamefont {Mayer}},\ }\bibfield  {title}
  {\enquote {\bibinfo {title} {Top {I}ncomes in {C}hile: {A} {H}istorical
  {P}erspective on {I}ncome {I}nequality, 1964--2017},}\ }\href@noop {}
  {\bibfield  {journal} {\bibinfo  {journal} {Review of Income and Wealth}\ }
  (\bibinfo {year} {2019})}\BibitemShut {NoStop}%
\bibitem [{\citenamefont {{OECD}}(2020)}]{OECD2020}%
  \BibitemOpen
  \bibfield  {author} {\bibinfo {author} {\bibnamefont {{OECD}}},\ }\href@noop
  {} {\enquote {\bibinfo {title} {{Income inequality (indicator)}},}\ }
  (\bibinfo {year} {2020}),\ \bibinfo {note} {{doi: 10.1787/459aa7f1-en
  (Accessed on 28 February 2020)}}\BibitemShut {NoStop}%
\bibitem [{\citenamefont {{{E}conomic {C}ommission for {L}atin {A}merica and
  the {C}aribbean (ECLAC)}}(209)}]{eclac2019}%
  \BibitemOpen
  \bibfield  {author} {\bibinfo {author} {\bibnamefont {{{E}conomic
  {C}ommission for {L}atin {A}merica and the {C}aribbean (ECLAC)}}},\
  }\href@noop {} {\enquote {\bibinfo {title} {Social {P}anorama of {L}atin
  {A}merica},}\ }\bibinfo {type} {Tech. Rep.}\ (\bibinfo  {institution} {United
  Nations},\ \bibinfo {year} {209})\BibitemShut {NoStop}%
\bibitem [{\citenamefont {Donoso}\ and\ \citenamefont
  {Bülow}(2017)}]{Donoso2017}%
  \BibitemOpen
  \bibfield  {author} {\bibinfo {author} {\bibfnamefont {S.}~\bibnamefont
  {Donoso}}\ and\ \bibinfo {author} {\bibfnamefont {M.}~\bibnamefont
  {Bülow}},\ }\href {\doibase 10.1057/978-1-137-60013-4} {\emph {\bibinfo
  {title} {Social movements in {C}hile: {O}rganization, trajectories, and
  political consequences}}}\ (\bibinfo  {publisher} {Palgrave Macmillan, New
  York},\ \bibinfo {year} {2017})\ pp.\ \bibinfo {pages} {1--286}\BibitemShut
  {NoStop}%
\bibitem [{Note1()}]{Note1}%
  \BibitemOpen
  \bibinfo {note} {Professor Claudia Sanhueza declared to BBC that ``This wave
  of protests may have been kickstarted by a rise in the price of metro
  tickets, but the resentment goes back further than that''. Moreover she
  ``singles out 2006 as a crucial year'' referring to the so-called ``Penguin
  Revolution'' led by high school students.
  https://www.bbc.com/news/world-latin-america-50151323.}\BibitemShut {Stop}%
\bibitem [{\citenamefont {Cabalin}(2012)}]{cabalin2012neoliberal}%
  \BibitemOpen
  \bibfield  {author} {\bibinfo {author} {\bibfnamefont {C.}~\bibnamefont
  {Cabalin}},\ }\bibfield  {title} {\enquote {\bibinfo {title} {Neoliberal
  education and student movements in {C}hile: {I}nequalities and malaise},}\
  }\href@noop {} {\bibfield  {journal} {\bibinfo  {journal} {Policy Futures in
  Education}\ }\textbf {\bibinfo {volume} {10}},\ \bibinfo {pages} {219--228}
  (\bibinfo {year} {2012})}\BibitemShut {NoStop}%
\bibitem [{Note2()}]{Note2}%
  \BibitemOpen
  \bibinfo {note} {The total population in Santiago is about 5.5 million and
  about 19 million in all Chile.}\BibitemShut {Stop}%
\bibitem [{\citenamefont {{{U}ndersecretary for {H}uman
  {R}ights}}(2020)}]{hhrr}%
  \BibitemOpen
  \bibfield  {author} {\bibinfo {author} {\bibnamefont {{{U}ndersecretary for
  {H}uman {R}ights}}},\ }\href@noop {} {\enquote {\bibinfo {title} {Informe
  actualizado de estado de situaci\'on desde el 19 de octubre},}\ } (\bibinfo
  {year} {2019-2020}),\ \bibinfo {note}
  {{http://ddhh.minjusticia.gob.cl/informacion-sobre-la-situacion-del-pais-desde-el-19-de-octubre}}\BibitemShut
  {NoStop}%
\bibitem [{\citenamefont {Epstein}(2002)}]{Epstein7243}%
  \BibitemOpen
  \bibfield  {author} {\bibinfo {author} {\bibfnamefont {J.~M.}\ \bibnamefont
  {Epstein}},\ }\bibfield  {title} {\enquote {\bibinfo {title} {Modeling civil
  violence: An agent-based computational approach},}\ }\href {\doibase
  10.1073/pnas.092080199} {\bibfield  {journal} {\bibinfo  {journal}
  {Proceedings of the National Academy of Sciences}\ }\textbf {\bibinfo
  {volume} {99}},\ \bibinfo {pages} {7243--7250} (\bibinfo {year}
  {2002})}\BibitemShut {NoStop}%
\bibitem [{\citenamefont {Pires}\ and\ \citenamefont
  {Crooks}(2017)}]{pires2017modeling}%
  \BibitemOpen
  \bibfield  {author} {\bibinfo {author} {\bibfnamefont {B.}~\bibnamefont
  {Pires}}\ and\ \bibinfo {author} {\bibfnamefont {A.}~\bibnamefont {Crooks}},\
  }\bibfield  {title} {\enquote {\bibinfo {title} {Modeling the emergence of
  riots: A geosimulation approach},}\ }\href {\doibase
  10.1016/j.compenvurbsys.2016.09.003} {\bibfield  {journal} {\bibinfo
  {journal} {Computers, Environment and Urban Systems}\ }\textbf {\bibinfo
  {volume} {61}},\ \bibinfo {pages} {66--80} (\bibinfo {year}
  {2017})}\BibitemShut {NoStop}%
\bibitem [{\citenamefont {Flamm}(1994)}]{flamm2005law}%
  \BibitemOpen
  \bibfield  {author} {\bibinfo {author} {\bibfnamefont {M.~W.}\ \bibnamefont
  {Flamm}},\ }\href@noop {} {\emph {\bibinfo {title} {Law and {O}rder: {S}treet
  {C}rime, {C}ivil {U}nrest, and the {C}risis of {L}iberalism in the 1960s}}}\
  (\bibinfo  {publisher} {Wiley-Blackwell},\ \bibinfo {year}
  {1994})\BibitemShut {NoStop}%
\bibitem [{\citenamefont {Walton}\ and\ \citenamefont
  {David}(2005)}]{walton1994free}%
  \BibitemOpen
  \bibfield  {author} {\bibinfo {author} {\bibfnamefont {J.~K.}\ \bibnamefont
  {Walton}}\ and\ \bibinfo {author} {\bibfnamefont {S.}~\bibnamefont {David}},\
  }\href@noop {} {\emph {\bibinfo {title} {Free {M}arkets and {F}ood {R}iots:
  {T}he {P}olitics of {G}lobal {A}djustment}}}\ (\bibinfo  {publisher}
  {Columbia University Press},\ \bibinfo {year} {2005})\BibitemShut {NoStop}%
\bibitem [{\citenamefont {Koselleck}\ and\ \citenamefont
  {Richter}(2006)}]{Koselleck}%
  \BibitemOpen
  \bibfield  {author} {\bibinfo {author} {\bibfnamefont {R.}~\bibnamefont
  {Koselleck}}\ and\ \bibinfo {author} {\bibfnamefont {M.~W.}\ \bibnamefont
  {Richter}},\ }\bibfield  {title} {\enquote {\bibinfo {title} {Crisis},}\
  }\href {http://www.jstor.org/stable/30141882} {\bibfield  {journal} {\bibinfo
   {journal} {Journal of the History of Ideas}\ }\textbf {\bibinfo {volume}
  {67}},\ \bibinfo {pages} {357--400} (\bibinfo {year} {2006})}\BibitemShut
  {NoStop}%
\bibitem [{\citenamefont {Beinhocker}(2006)}]{beinhocker2006origin}%
  \BibitemOpen
  \bibfield  {author} {\bibinfo {author} {\bibfnamefont {E.}~\bibnamefont
  {Beinhocker}},\ }\href {https://books.google.cl/books?id=eUoolrxSFy0C} {\emph
  {\bibinfo {title} {The {O}rigin of {W}ealth: {E}volution, {C}omplexity, and
  the {R}adical {R}emaking of {E}conomics}}}\ (\bibinfo  {publisher} {Harvard
  Business School Press},\ \bibinfo {year} {2006})\BibitemShut {NoStop}%
\bibitem [{\citenamefont {Arnold}(1984)}]{Catastrophe}%
  \BibitemOpen
  \bibfield  {author} {\bibinfo {author} {\bibfnamefont {V.}~\bibnamefont
  {Arnold}},\ }\href@noop {} {\emph {\bibinfo {title} {{C}atastrophe
  {T}heory}}}\ (\bibinfo  {publisher} {Springer, Heidelberg},\ \bibinfo {year}
  {1984})\BibitemShut {NoStop}%
\bibitem [{\citenamefont {Zeeman}(1976)}]{Zeeman}%
  \BibitemOpen
  \bibfield  {author} {\bibinfo {author} {\bibfnamefont {E.~C.}\ \bibnamefont
  {Zeeman}},\ }\bibfield  {title} {\enquote {\bibinfo {title} {Catastrophe
  theory},}\ }\href {http://www.jstor.org/stable/24950329} {\bibfield
  {journal} {\bibinfo  {journal} {Scientific American}\ }\textbf {\bibinfo
  {volume} {234}},\ \bibinfo {pages} {65--83} (\bibinfo {year}
  {1976})}\BibitemShut {NoStop}%
\bibitem [{\citenamefont {Scherman}, \citenamefont {Arriagada},\ and\
  \citenamefont {Valenzuela}(2015)}]{scherman2015student}%
  \BibitemOpen
  \bibfield  {author} {\bibinfo {author} {\bibfnamefont {A.}~\bibnamefont
  {Scherman}}, \bibinfo {author} {\bibfnamefont {A.}~\bibnamefont {Arriagada}},
  \ and\ \bibinfo {author} {\bibfnamefont {S.}~\bibnamefont {Valenzuela}},\
  }\bibfield  {title} {\enquote {\bibinfo {title} {Student and environmental
  protests in chile: The role of social media},}\ }\href@noop {} {\bibfield
  {journal} {\bibinfo  {journal} {Politics}\ }\textbf {\bibinfo {volume}
  {35}},\ \bibinfo {pages} {151--171} (\bibinfo {year} {2015})}\BibitemShut
  {NoStop}%
\bibitem [{\citenamefont {Boulianne}(2015)}]{boulianne2015social}%
  \BibitemOpen
  \bibfield  {author} {\bibinfo {author} {\bibfnamefont {S.}~\bibnamefont
  {Boulianne}},\ }\bibfield  {title} {\enquote {\bibinfo {title} {Social media
  use and participation: A meta-analysis of current research},}\ }\href@noop {}
  {\bibfield  {journal} {\bibinfo  {journal} {Information, communication \&
  society}\ }\textbf {\bibinfo {volume} {18}},\ \bibinfo {pages} {524--538}
  (\bibinfo {year} {2015})}\BibitemShut {NoStop}%
\bibitem [{\citenamefont {Weeks}, \citenamefont {Ard{\`e}vol-Abreu},\ and\
  \citenamefont {Gil~de Z{\'u}{\~n}iga}(2017)}]{weeks2017online}%
  \BibitemOpen
  \bibfield  {author} {\bibinfo {author} {\bibfnamefont {B.~E.}\ \bibnamefont
  {Weeks}}, \bibinfo {author} {\bibfnamefont {A.}~\bibnamefont
  {Ard{\`e}vol-Abreu}}, \ and\ \bibinfo {author} {\bibfnamefont
  {H.}~\bibnamefont {Gil~de Z{\'u}{\~n}iga}},\ }\bibfield  {title} {\enquote
  {\bibinfo {title} {Online influence? social media use, opinion leadership,
  and political persuasion},}\ }\href@noop {} {\bibfield  {journal} {\bibinfo
  {journal} {International Journal of Public Opinion Research}\ }\textbf
  {\bibinfo {volume} {29}},\ \bibinfo {pages} {214--239} (\bibinfo {year}
  {2017})}\BibitemShut {NoStop}%
\bibitem [{\citenamefont {Valenzuela}\ \emph {et~al.}(2016)\citenamefont
  {Valenzuela}, \citenamefont {Somma}, \citenamefont {Scherman},\ and\
  \citenamefont {Arriagada}}]{valenzuela2016social}%
  \BibitemOpen
  \bibfield  {author} {\bibinfo {author} {\bibfnamefont {S.}~\bibnamefont
  {Valenzuela}}, \bibinfo {author} {\bibfnamefont {N.~M.}\ \bibnamefont
  {Somma}}, \bibinfo {author} {\bibfnamefont {A.}~\bibnamefont {Scherman}}, \
  and\ \bibinfo {author} {\bibfnamefont {A.}~\bibnamefont {Arriagada}},\
  }\bibfield  {title} {\enquote {\bibinfo {title} {Social media in latin
  america: deepening or bridging gaps in protest participation?}}\ }\href@noop
  {} {\bibfield  {journal} {\bibinfo  {journal} {Online Information Review}\ }
  (\bibinfo {year} {2016})}\BibitemShut {NoStop}%
\bibitem [{\citenamefont {Kermack}, \citenamefont {McKendrick},\ and\
  \citenamefont {Walker}(1927)}]{Kermack}%
  \BibitemOpen
  \bibfield  {author} {\bibinfo {author} {\bibfnamefont {W.~O.}\ \bibnamefont
  {Kermack}}, \bibinfo {author} {\bibfnamefont {A.~G.}\ \bibnamefont
  {McKendrick}}, \ and\ \bibinfo {author} {\bibfnamefont {G.~T.}\ \bibnamefont
  {Walker}},\ }\bibfield  {title} {\enquote {\bibinfo {title} {A contribution
  to the mathematical theory of epidemics},}\ }\href {\doibase
  10.1098/rspa.1927.0118} {\bibfield  {journal} {\bibinfo  {journal}
  {Proceedings of the Royal Society of London. Series A, Containing Papers of a
  Mathematical and Physical Character}\ }\textbf {\bibinfo {volume} {115}},\
  \bibinfo {pages} {700--721} (\bibinfo {year} {1927})},\ \Eprint
  {http://arxiv.org/abs/https://royalsocietypublishing.org/doi/pdf/10.1098/rspa.1927.0118}
  {https://royalsocietypublishing.org/doi/pdf/10.1098/rspa.1927.0118}
  \BibitemShut {NoStop}%
\bibitem [{\citenamefont {Heesterbeek}(2002)}]{Heesterbeek2002}%
  \BibitemOpen
  \bibfield  {author} {\bibinfo {author} {\bibfnamefont {J.~A.~P.}\
  \bibnamefont {Heesterbeek}},\ }\bibfield  {title} {\enquote {\bibinfo {title}
  {A brief history of {R0} and a recipe for its calculation},}\ }\href@noop {}
  {\bibfield  {journal} {\bibinfo  {journal} {Acta biotheoretica}\ }\textbf
  {\bibinfo {volume} {50}},\ \bibinfo {pages} {189--204} (\bibinfo {year}
  {2002})}\BibitemShut {NoStop}%
\bibitem [{\citenamefont {Brauer}, \citenamefont {Castillo-Chavez},\ and\
  \citenamefont {Castillo-Chavez}(2012)}]{Brauer2012}%
  \BibitemOpen
  \bibfield  {author} {\bibinfo {author} {\bibfnamefont {F.}~\bibnamefont
  {Brauer}}, \bibinfo {author} {\bibfnamefont {C.}~\bibnamefont
  {Castillo-Chavez}}, \ and\ \bibinfo {author} {\bibfnamefont {C.}~\bibnamefont
  {Castillo-Chavez}},\ }\href@noop {} {\emph {\bibinfo {title} {Mathematical
  models in population biology and epidemiology}}},\ Vol.~\bibinfo {volume}
  {2}\ (\bibinfo  {publisher} {Springer},\ \bibinfo {year} {2012})\BibitemShut
  {NoStop}%
\bibitem [{\citenamefont {Arnold}\ and\ \citenamefont
  {Avez}(1968)}]{ArnoldAvez}%
  \BibitemOpen
  \bibfield  {author} {\bibinfo {author} {\bibfnamefont {V.}~\bibnamefont
  {Arnold}}\ and\ \bibinfo {author} {\bibfnamefont {A.}~\bibnamefont {Avez}},\
  }\href@noop {} {\emph {\bibinfo {title} {Ergodic Problems of Classical
  Mechanics}}}\ (\bibinfo  {publisher} {(The Mathematical Physics Monograph
  Series) New York/Amsterdam W. A. Benjamin, Inc.},\ \bibinfo {year}
  {1968})\BibitemShut {NoStop}%
\bibitem [{\citenamefont {Kloeden}\ and\ \citenamefont
  {Platten}(1995)}]{EulerMayurama}%
  \BibitemOpen
  \bibfield  {author} {\bibinfo {author} {\bibfnamefont {P.}~\bibnamefont
  {Kloeden}}\ and\ \bibinfo {author} {\bibfnamefont {E.}~\bibnamefont
  {Platten}},\ }\href@noop {} {\emph {\bibinfo {title} {Numerical {S}olution of
  {S}tochastic {D}ifferential {E}quations}}}\ (\bibinfo  {publisher} {Springer,
  Heidelberg},\ \bibinfo {year} {1995})\BibitemShut {NoStop}%
\bibitem [{\citenamefont {Bak}, \citenamefont {Tang},\ and\ \citenamefont
  {Wiesenfeld}(1987)}]{soc}%
  \BibitemOpen
  \bibfield  {author} {\bibinfo {author} {\bibfnamefont {P.}~\bibnamefont
  {Bak}}, \bibinfo {author} {\bibfnamefont {C.}~\bibnamefont {Tang}}, \ and\
  \bibinfo {author} {\bibfnamefont {K.}~\bibnamefont {Wiesenfeld}},\ }\bibfield
   {title} {\enquote {\bibinfo {title} {Self-organized criticality: An
  explanation of the 1/f noise},}\ }\href {\doibase 10.1103/PhysRevLett.59.381}
  {\bibfield  {journal} {\bibinfo  {journal} {Phys. Rev. Lett.}\ }\textbf
  {\bibinfo {volume} {59}},\ \bibinfo {pages} {381--384} (\bibinfo {year}
  {1987})}\BibitemShut {NoStop}%
\bibitem [{\citenamefont {Bak}(1996)}]{bak}%
  \BibitemOpen
  \bibfield  {author} {\bibinfo {author} {\bibfnamefont {P.}~\bibnamefont
  {Bak}},\ }\href@noop {} {\emph {\bibinfo {title} {How {N}ature {W}orks: {T}he
  {S}cience of {S}elf-{O}rganized {C}riticality}}}\ (\bibinfo  {publisher}
  {Springer, Heidelberg},\ \bibinfo {year} {1996})\BibitemShut {NoStop}%
\bibitem [{\citenamefont {Rhodes}, \citenamefont {Jensen},\ and\ \citenamefont
  {Anderson}(1997)}]{Rhodes1997}%
  \BibitemOpen
  \bibfield  {author} {\bibinfo {author} {\bibfnamefont {C.~J.}\ \bibnamefont
  {Rhodes}}, \bibinfo {author} {\bibfnamefont {H.~J.}\ \bibnamefont {Jensen}},
  \ and\ \bibinfo {author} {\bibfnamefont {R.~M.}\ \bibnamefont {Anderson}},\
  }\bibfield  {title} {\enquote {\bibinfo {title} {On the critical behaviour of
  simple epidemics},}\ }\href {\doibase 10.1098/rspb.1997.0228} {\bibfield
  {journal} {\bibinfo  {journal} {Proceedings of the Royal Society of London.
  Series B: Biological Sciences}\ }\textbf {\bibinfo {volume} {264}},\ \bibinfo
  {pages} {1639--1646} (\bibinfo {year} {1997})},\ \Eprint
  {http://arxiv.org/abs/https://royalsocietypublishing.org/doi/pdf/10.1098/rspb.1997.0228}
  {https://royalsocietypublishing.org/doi/pdf/10.1098/rspb.1997.0228}
  \BibitemShut {NoStop}%
\bibitem [{Note3()}]{Note3}%
  \BibitemOpen
  \bibinfo {note} {``Estamos en guerra contra un enemigo poderoso, implacable,
  que no respeta a nada ni a nadie y que est\'a dispuesto a usar la violencia y
  la delincuencia sin ning\'un l\'imite''. President Sebasti\'an Pi\~nera in
  press conference on Sunday Oct. 20$^{\protect \rm th}$, 2019.}\BibitemShut
  {Stop}%
\end{thebibliography}
\end{document}